\documentclass[onecolumn, a4size, 12pt]{IEEEtran}
\usepackage{amsmath}
\usepackage{amssymb}
\usepackage{amsfonts}
\usepackage{graphicx}
\usepackage{epsfig}
\usepackage{subfigure}
\usepackage{psfrag}

\title{On Multiuser Power Region of Fading Multiple-Access Channel with Multiple Antennas\footnote{Submitted to
IEEE Transactions on Information Theory, September 2006, revised
April 2008. This work was supported partially by NSF grant
CNS-0427711. Part of this paper has been presented at IEEE
International Symposium on Information Theory (ISIT), Seattle, July
2006. Please address all correspondence to Rui Zhang.}}

\author{Rui Zhang, Mehdi Mohseni, and John M. Cioffi \footnote{Rui Zhang is with the Institute for Infocomm Research, A*STAR, Singapore.
(e-mail: rzhang@i2r.a-star.edu.sg)} \footnote{Mehdi Moshseni is with
the ASSIA, Inc. (e-mail:mmohseni@assia-inc.com)} \footnote{John M.
Cioffi is with the Department of Electrical Engineering, Stanford
University. (e-mail: cioffi@stanford.edu) }}

\begin{document}
\maketitle \maketitle \thispagestyle{empty}

\begin{abstract}
This paper is concerned with the fading multiple-input
multiple-output multiple-access channel (MIMO-MAC) with multiple
receive antennas at the base station (BS) and multiple transmit
antennas at each mobile terminal (MT). Two multiple-access
techniques are considered for scheduling transmissions from each MT
to the BS at the same frequency, which are {\it space-division
multiple-access} (SDMA) and {\it time-division multiple-access}
(TDMA). For SDMA, all MTs transmit simultaneously to the BS and
their individual signals are resolved at the BS via multiple receive
antennas while for TDMA, each MT transmits independently to the BS
during mutually orthogonal time slots. It is assumed that the
channel-state information (CSI) of the fading channel from each MT
to the BS is {\it unknown} at each MT transmitter, but is perfectly
{\it known} at the BS receiver. Thereby, the BS can acquire the
long-term channel-distribution information (CDI) for each MT. This
paper extends the well-known {\it transmit-covariance feedback}
scheme for the point-to-point fading MIMO channel to the fading
MIMO-MAC, whereby the BS jointly optimizes the transmit signal
covariance matrices for all MTs based on their CDI, and then sends
each transmit covariance matrix back to the corresponding MT via a
feedback channel. The main goal of this paper is to characterize the
so-called {\it multiuser power region} under the multiuser
transmit-covariance feedback scheme for both SDMA and TDMA. The
power region is defined as the constitution of all user transmit
power-tuples that can achieve reliable transmissions for a given set
of user target rates. Simulation results show that SDMA can achieve
substantial power savings over TDMA for the fading MIMO-MAC, even
when the number of antennas at the BS is equal to that at each MT.
Moreover, this paper demonstrates the usefulness of the multiuser
power region for maintaining {\it proportionally-fair} power
consumption among the MTs.
\end{abstract}


\setlength{\baselineskip}{1.3\baselineskip}
\newtheorem{claim}{Claim}
\newtheorem{guess}{Conjecture}
\newtheorem{definition}{Definition}
\newtheorem{fact}{Fact}
\newtheorem{assumption}{Assumption}
\newtheorem{theorem}{Theorem}
\newtheorem{lemma}{\underline{Lemma}}
\newtheorem{ctheorem}{Corrected Theorem}
\newtheorem{corollary}{Corollary}
\newtheorem{proposition}{Proposition}
\newtheorem{example}{Example}
\newtheorem{remark}{Remark}[section]
\newtheorem{problem}{\underline{Problem}}
\newtheorem{algorithm}{\underline{Algorithm}}
\newcommand{\mv}[1]{\mbox{\boldmath{$ #1 $}}}

\newpage

\section{Introduction}
Transmission through multiple transmit and multiple receive
antennas, or the so-called multiple-input multiple-output (MIMO)
technology, is known as an efficient means for providing enormous
information rates in rich-scattering mobile environments
\cite{Telatar99}-\cite{Raleigh98}. Characterization of the fading
MIMO channel capacity limits, under various assumptions on the
transmitter-side and receiver-side channel-state information (CSI)
and channel-distribution information (CDI), has motivated a great
deal of valuable scholarly work (e.g., \cite{Goldsmith03} and
references therein). In particular, the case where the CSI is
perfectly {\it known} at the receiver but {\it unknown} at the
transmitter has drawn much interest due to its validity in many
practical situations. This is because the presumption of perfect
receiver-side CSI is usually reasonable for wireless channels where
the receiver can locally estimate the fading channel, while the
complete CSI feedback from the receiver to the transmitter is
difficult or even impossible. Consequently, many schemes that
exploit various forms of {\it partial} channel feedback have been
proposed in literature. Among others, the {\it transmit-covariance
feedback} scheme is known to be capable of achieving data rates
close to the fading MIMO channel ergodic capacity when the channel
CDI exhibits some long-term consistent statistical properties, e.g.,
constant channel mean and/or constant channel covariance matrix
\cite{Madhow01}-\cite{Boche04a}. In the transmit-covariance feedback
scheme, the receiver determines the transmit signal covariance
matrix based on the CDI, and then sends it back to the transmitter
through a feedback channel. In \cite{Madhow01}-\cite{Boche04a}, the
feedback transmit covariance matrix for optimizing the channel
ergodic capacity, and the conditions under which the beamforming --
the transmit covariance matrix has rank one -- is optimal, have been
established for the single-user fading channel. In this scheme, the
transmit covariance matrix is fixed as long as the CDI is not
changed. Therefore, this scheme requires much less feedback
complexity and is also more robust to the delay of the feedback
channel compared to other partial channel feedback schemes based on
the instantaneous MIMO channel realizations (e.g., \cite{Heath04},
\cite{Zhang07} and references therein).

This paper considers the fading MAC with additive white Gaussian
noise (AWGN) at the receiver, and assumes that the CSI from each
mobile terminal (MT) to the base station (BS) is unknown at each MT
transmitter, but is perfectly known at the BS receiver. Thus, the BS
can acquire the channel CDI for each MT. This paper extends the
transmit-covariance feedback scheme for the single-user fading MIMO
channel to the fading MIMO multiple-access channel (MIMO-MAC) where
multiple antennas are used by the BS and possibly by each MT. Two
multiple-access techniques are considered for scheduling
transmissions from each MT to the BS at the same frequency: {\it
space-division multiple-access} (SDMA) and {\it time-division
multiple-access} (TDMA). For SDMA, all MTs transmit simultaneously
to the BS and their individual signals are decoded jointly at the BS
while for TDMA, each MT transmits independently to the BS during
mutually orthogonal time slots and thus only single-user decoding is
needed. The multiuser transmit-covariance feedback scheme is then
described as follows. For SDMA, the BS first jointly optimizes the
transmit signal covariance matrices for all MTs, based on the
multiuser CDI as well as the rate requirement and the power budget
of each MT, and then sends them back to each MT for transmission.
This scheme has also been considered in \cite{Syed01},
\cite{Ulukus05} for characterizing the capacity region and
establishing the conditions for the optimality of beamforming for
the fading MIMO-MAC, respectively. In contrast, for TDMA, the BS
jointly optimizes the duration of transmission time slot for each MT
along with their transmit covariance matrices. These optimized
values are then sent back to each corresponding MT via the feedback
channel.

This paper studies the information-theoretic limits of the fading
MIMO-MAC under the multiuser transmit-covariance feedback scheme
when either SDMA or TDMA is employed. Two commonly adopted means to
measure the information-theoretic limits of multiuser channels are
the {\it capacity region} and the {\it power region}. The capacity
region is defined as the constitution of all achievable rate-tuples
for the users given their individual power constraints, while the
power region consists of all possible power-tuples for the users
under which a given rate-tuple is achievable. This paper is mainly
concerned with the characterization of the multiuser power region.
Our motivations are justified as follows:

First, characterization of the power region for the fading MIMO-MAC
is a challenging problem. Considering initially the case of SDMA,
the capacity region of a {\it deterministic} (no fading) Gaussian
MAC with a single transmit and a single receive antenna (SISO-MAC)
has the well-known {\it polymatroid} structure \cite{Tse98a}, which
also holds for the fading MIMO-MAC. On the other hand, the power
region of a {\it deterministic} SISO-MAC is known to have a {\it
contra-polymatroid} structure \cite{Tse98b}. The polymatroid and the
contra-polymatroid structures can be utilized to reduce
significantly the computational complexity of finding the boundary
points of the capacity region and the power region, respectively
\cite{Tse98a}, \cite{Tse98b}. However, the contra-polymatroid
structure is non-existent for the power region when the channel
exhibits fading \cite{Mecking00} and/or the BS uses multiple
antennas \cite{Muller99}.\footnote{More discussions on this aspect
are postponed to Section \ref{subsec:weighted sum power}.} As a
result, characterization of the power region for the fading MIMO-MAC
under SDMA is yet fully understood in literature. On the other hand,
for TDMA, given the duration of each MT transmission time slot
(e.g., equal time-slot durations for all MTs in the conventional
TDMA), the BS only needs to optimize the transmit covariance
matrices for the MTs independently such that their individual
transmission powers are minimized for supporting their own target
rates. However, with time-slot duration adjustment for each MT, the
BS now needs to consider the more challenging problem of jointly
optimizing the time-slot durations and the transmit covariance
matrices for all MTs. This joint optimization has been less studied
in literature.

Secondly, characterization of the power region can potentially
provide very useful insights on the resource allocation problem for
wireless networks, e.g., the wireless cellular network. In many
practical situations, each BS in the cellular network controls the
transmit power of each MT in its cell such that each individual MT
rate demand -- transmission quality-of-service (QoS) -- is satisfied
(e.g., \cite{Yates95}, \cite{Liu98}). Power control can be
beneficial in many aspects, e.g., to maintain fair power consumption
among MTs, to tailor for each MT's peak-power constraint, and to
mitigate the co-channel interference between multiple cells
operating at the same frequency. By exploiting the multiuser power
region, the minimum power consumption in the network can be achieved
under practical transmission constraints.

The main contributions of this paper are summarized as follows:
\begin{itemize}
\item For both SDMA and TDMA, the paper presents efficient algorithms for
characterizing each boundary point of the power region for the
fading MIMO-MAC. The developed algorithms are based on a Lagrange
primal-dual approach that exploits a novel dual relationship between
the power region and the corresponding capacity region for the
fading MIMO-MAC. For SDMA, the proposed algorithm determines jointly
the optimal transmit covariance matrices for all MTs as well as
their optimal decoding order at the receiver. For TDMA, all MT
transmit covariance matrices along with their assigned variable
time-slot durations are jointly optimized.

\item In addition to the conventional way to characterize the
boundary of the power region by solving a sequence of user {\it
weighted} sum-power minimization problems subject to fixed user rate
constraints, this paper proposes an alternative means for such
characterization by considering the user sum-power minimization
problem under different {\it user power-profile} constraints, where
the user power-profile regulates the power consumption of users
under some given proportional fairness.
\end{itemize}

The remainder of this paper is organized as follows. Section
\ref{sec:system model} introduces the system model for the fading
MIMO-MAC and describes the proposed multiuser transmit-covariance
feedback scheme under SDMA and TDMA. Section \ref{sec:power region}
provides the definition of the power region for the fading MIMO-MAC,
together with two problem formulations for characterization of the
power region, based on the user weighted sum-power minimization and
the user power-profile vector, respectively. Section \ref{sec:SDMA}
and Section \ref{sec:TDMA} study the power region in the case of
SDMA and TDMA, respectively, and present efficient algorithms for
characterizing the power region in each case. Section
\ref{sec:simulations} provides numerical results to verify the
usefulness of the proposed algorithms under realistic channel
parameters. Finally, Section \ref{sec:conclusions} concludes the
paper.

{\it Notations}: Scalar signals are denoted by lower-case letters,
e.g., $x, y$. Bold-face lower-case letters are used for vector
signals, e.g., $\mv{x}, \mv{y}$, and bold-face upper-case letters
for matrices, e.g., $\mv{S}$, $\mv{M}$. $|\mv{S}|$ denotes the
determinant, $\mv{S}^{-1}$ the inverse and $\mathtt{Tr}(\mv{S})$ the
trace of a square matrix $\mv{S}$. For any general matrix $\mv{M}$,
$\mv{M}^{T}$ and $\mv{M}^{\dag}$ denote its transpose and conjugate
transpose, respectively. $\mv{I}$ denotes the identity matrix.
$\mathbb{E}[\cdot]$ denotes statistical expectation. $\mathbb{C}^{x
\times y}$ denotes the space of $x\times y$ matrices with complex
entries. $\mathbb{R}^M$ denotes the $M$-dimensional real Euclidean
space and $\mathbb{R}^M_{+}$ is the nonnegative orthant. The
distribution of a circular symmetric complex Gaussian (CSCG) vector
with the mean vector $\mv{x}$, and the covariance matrix
$\mv{\Sigma}$ is denoted by $\mathcal{CN}(\mv{x},\mv{\Sigma})$, and
$\sim$ means ``distributed as.'' The sign $\succeq$ denotes the
generalized inequality \cite{Boydbook} and for a square matrix
$\mv{S}$, $\mv{S} \succeq 0$ means that $\mv{S}$ is positive
semi-definite. $\min(x,y)$ denotes the minimum between two real
numbers $x$ and $y$.

\section{System Model} \label{sec:system model}
This paper considers a narrow-band flat fading MIMO-MAC with $r$
receive antennas at the BS and $K$ MTs equipped with
$t_1,\ldots,t_K$ transmit antennas, respectively. All MTs transmit
synchronously to the BS by sharing a common frequency band. It is
assumed that the space of fading states is continuous and infinite,
and the fading process is stationary and ergodic. Under the
assumption that the transmitted symbol period is equal to the
inverse of the common transmission bandwidth for all MTs, at each
fading state $\nu$, the equivalent discrete-time MAC is given by
\begin{eqnarray} \label{eq:MIMO MAC}
\mv{y}=\sum_{k=1}^{K} \mv{H}_k(\nu)\mv{x}_k + \mv{z},
\end{eqnarray}
where $\mv{y}\in\mathbb{C}^{r\times 1}$ denotes the received
signal vector,  $\mv{x}_k\in\mathbb{C}^{t_k\times 1}$ and
$\mv{H}_k(\nu)\in\mathbb{C}^{r\times t_k}$ denote, respectively,
the transmitted signal vector and the channel matrix of MT $k$,
$k=1,\ldots,K$; $\mv{z}\in\mathbb{C}^{r\times 1}$ denotes the
vector of additive noise at the receiver, and it is assumed that
$\mv{z}\sim \mathcal{CN}(0,\mv{I})$.

\begin{figure}
\psfrag{a}{Base Station} \psfrag{b}{MT 1} \psfrag{c}{MT 2}
\psfrag{d}{MT K} \psfrag{e}{$\mv{H}_1(\nu)$}
\psfrag{f}{$\mv{H}_2(\nu)$} \psfrag{g}{$\mv{H}_K(\nu)$}
\psfrag{h}{{\it Feedback Channel}} \psfrag{i}{$\mv{S}_1$}
\psfrag{j}{$\mv{S}_2$} \psfrag{k}{$\mv{S}_K$}
\begin{center}
\scalebox{1.2}{\includegraphics*[41pt,461pt][442pt,721pt]{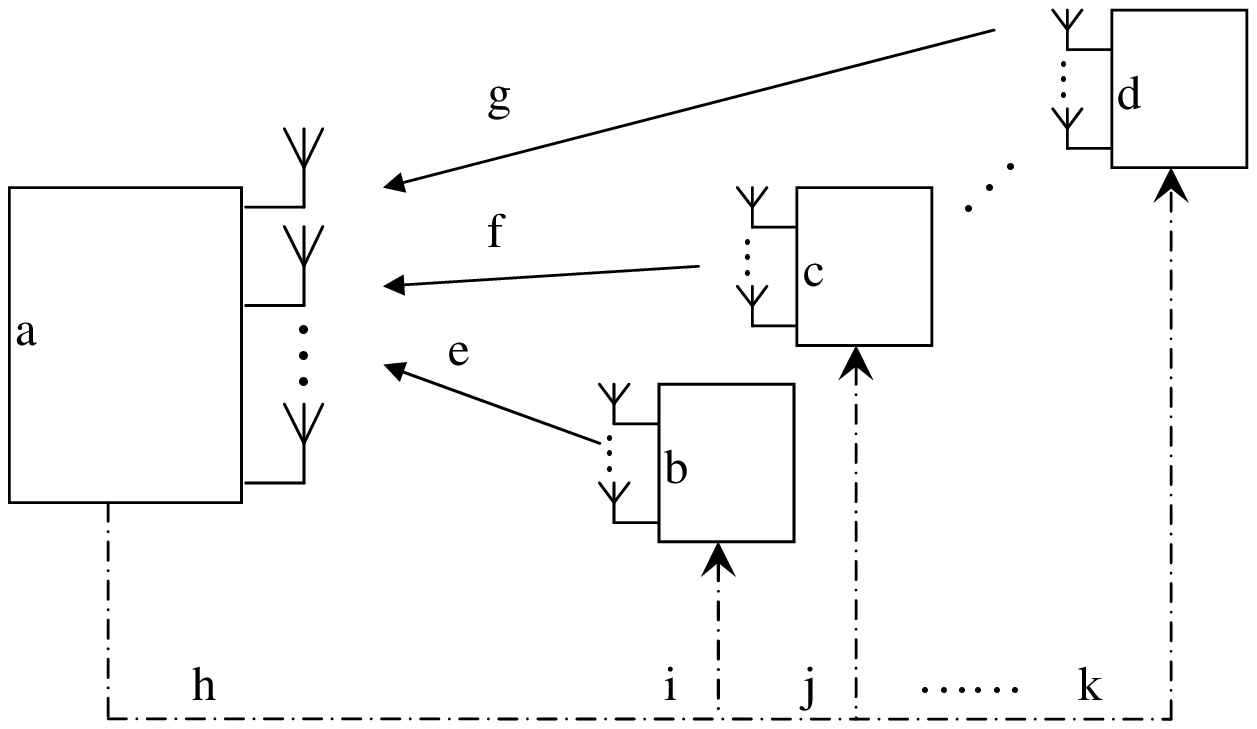}}
\end{center}
\caption{The fading MIMO-MAC with the multiuser
transmit-covariance feedback.} \label{fig:MAC model}
\end{figure}

This channel model also assumes that the CSI is perfectly known at
the BS but is unknown at each MT. With the CSI available, the BS can
acquire the long-term CSI statistics (or equivalently, the CDI) of
each MT. Based on the multiuser CDI, the BS determines the transmit
signal covariance matrices for all MTs jointly according to their
individual rate requirement and power budget, and then sends them
back to each MT for transmission. This paper refers to this scheme
as {\it multiuser transmit-covariance feedback}. Let the transmit
covariance matrix of MT $k$ be $\mv{S}_k\triangleq\mathbb{E}
[\mv{x}_k\mv{x}_k^{\dag}]$, where the expectation is taken over the
code-book and $\mv{S}_k\succeq 0$. $\mv{S}_k$ is assumed to be {\it
fixed} for all fading states $\nu$. Fig. \ref{fig:MAC model}
illustrates the system model considered in this paper. Since this
paper is concerned with the information-theoretic limits of a
Gaussian MAC, the optimal Gaussian code-book is assumed for each MT,
i.e., $\mv{x}_k\sim\mathcal{CN}(0,\mv{S}_k), \forall k$. The
transmit covariance matrix of MT $k$ can be expressed by its
eigenvalue decomposition as
\begin{equation}
\mv{S}_k=\mv{V}_k\mv{\Sigma}_k\mv{V}_k^{\dag}.
\end{equation}
$\mv{V}_k\in\mathbb{C}^{t_k\times d_k}$ is known as the {\it
precoding matrix} where $\mv{V}_k^{\dag}\mv{V}_k=\mv{I}$, and
$d_k\leq\min(t_k,r)$. $d_k$ is usually referred to as the {\it
spatial multiplexing gain} as it measures the number of degrees of
transmission freedom (or equivalently, the number of data streams)
in the spatial domain. If $d_k$ is equal to one, the associated
transmission scheme is usually referred to as {\it beamforming}.
$\mv{\Sigma}_k$ is a $d_k \times d_k$ diagonal matrix with positive
diagonal elements that provide the {\it power-loading} to the
corresponding transmitted data streams. The transmitter of each MT
can be implemented as the cascade of the following operations:
encoding the information bits by the optimal Gaussian code-book,
interleaving coded symbols randomly into each data stream,
power-loading and then jointly precoding all data streams. Next, two
multiple-access techniques considered in this paper are illustrated,
namely, SDMA and TDMA.

For SDMA, all MTs transmit simultaneously to the BS. In this paper,
it is assumed that the BS receiver uses the optimal
(capacity-achieving) multiuser detection. For a fixed set of
$\{\mv{S}_k\}$, $k=1,\ldots,K$, all the rate-tuples in the set,
$\mathcal{C}_{\mathtt{SDMA}}(\{\mv{S}_k\})$, defined below, are
achievable (e.g., \cite{Syed01}, \cite{Ulukus05}):
\begin{eqnarray}\label{eq:definition C SDMA}
\mathcal{C}_{\mathtt{SDMA}}(\{\mv{S}_k\}) = \left \{\mv{r}\in
\mathbb{R}_{+}^K: \sum_{k\in \mathcal{J}}r_k \leq
\mathbb{E}_{\nu}\left[\frac{1}{2} \log \left| \sum_{k\in
\mathcal{J}}\mv{H}_k(\nu)\mv{S}_k\mv{H}_k^{\dag}(\nu)+\mv{I}\right|\right]
\ \forall \mathcal{J}\subseteq \{1,\ldots,K\} \right\}.
\end{eqnarray}
The code-book of MT $k$ should satisfy $\mathtt{Tr}(\mv{S}_k)=p_k$,
where $\mv{p}=(p_1,\ldots,p_K)\in\mathbb{R}_{+}^K$ denotes the
vector of average transmit powers for the MTs.

On the other hand, for TDMA, the BS divides the total transmission
time into multiple transmission blocks of equal duration $T$. Each
transmission block is then further divided into $K$ non-overlapping
time slots assigned to the MTs. These time slots are assumed fixed
over all blocks. During the time slot of MT $k$, only this MT
communicates with the BS and other MTs are silent, i.e.,
$\mv{x}_{k'}=0, \forall k'\neq k$ in (\ref{eq:MIMO MAC}). Let
$\tau_kT$ denote the time-slot duration assigned to MT $k$, where
$0\leq\tau_k\leq 1, \forall k$ and $\sum_{k=1}^{K}\tau_k=1$. The BS
determines jointly the slot duration for each MT and their transmit
covariance matrices, and then sends them back to each MT. For a
fixed set of $\{\tau_k\}$ and $\{\mv{S}_k\}$, $k=1,\ldots,K$, each
MT transmits over a single-user fading MIMO channel studied in e.g.,
\cite{Telatar99}, \cite{Foschini98}, and thus the following
rate-tuples in the set, $\mathcal{C}_{\mathtt{TDMA}}(\{\tau_k\},
\{\mv{S}_k\})$, defined below are achievable:
\begin{eqnarray}\label{eq:definition C TDMA}
\mathcal{C}_{\mathtt{TDMA}}(\{\tau_k\},\{\mv{S}_k\}) = \left
\{\mv{r}\in \mathbb{R}_{+}^K: r_k \leq
\tau_k\mathbb{E}_{\nu}\left[\frac{1}{2} \log \left|
\mv{H}_k(\nu)\frac{\mv{S}_k}{\tau_k}\mv{H}_k^{\dag}(\nu)+\mv{I}\right|\right]
\ \forall k\in \{1,\ldots,K\} \right\}.
\end{eqnarray}
And, again, $\mv{p}=(p_1,\ldots,p_K)\in\mathbb{R}_{+}^K$ denotes the
average transmit powers for the MTs. Notice that for TDMA, the
actual transmit power for MT $k$ during its assigned time-slot
duration $\tau_kT$ is $\mathtt{Tr}\left(\mv{S}_k\right)/ \tau_k$,
but the average transmit power $p_k$ over each block duration $T$ is
$\mathtt{Tr}(\mv{S}_k)$, the same as SDMA.

\begin{remark}
In this work, for both SDMA and TDMA, we consider each user's
achievable rate in the ``ergodic'' sense, i.e., averaged over all
ergodic fading states. In the case of fast-fading channel, the
resultant ergodic capacity can be achievable by assigning each MT a
constant-rate code-book for which the codeword length is
sufficiently long so as to exploit the ergodicity of the channel. In
contrast, in the case of slow-fading channel, each MT's codeword
might not be able to span over all possible fading states because of
practical transmission delay constraint. However, if each MT uses
multiple code-books with variable rates, the BS, based on the
instantaneous channel, can determine the transmission rate of each
MT and then sends back the corresponding code-book index to each MT
for transmission. As in the fast-fading case, the same ergodic
capacity (sometimes known as the expected capacity) can be achieved
for each MT via time-averaging its transmission rates over different
fading states.
\end{remark}

\section{Power Region for Fading MIMO-MAC: Definitions and Characterizations} \label{sec:power region}
\begin{figure}
\psfrag{a}{$p_1$} \psfrag{b}{$p_2$} \psfrag{c}{WSPmin}
\psfrag{d}{$\mv{\alpha}$} \psfrag{e}{Power Profile}
\psfrag{f}{$\mv{\lambda}^T\mv{p}$}
\begin{center}
\scalebox{1.0}{\includegraphics*[33pt,567pt][256pt,788pt]{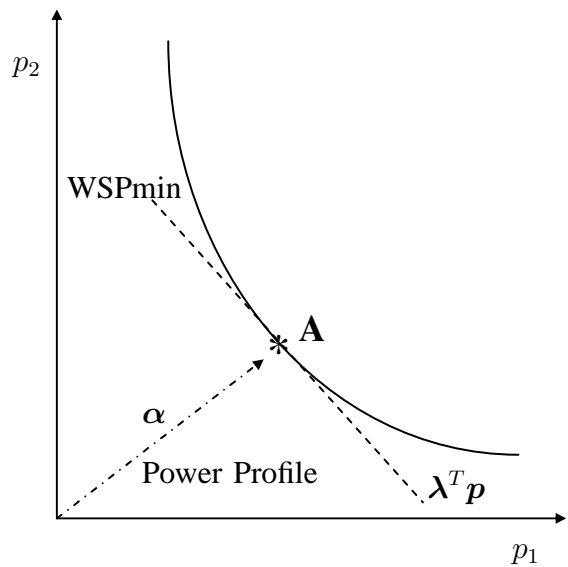}}
\end{center}
\caption{Characterization of power region via WSPmin or
power-profile vector.} \label{fig:power region defined}
\end{figure}

The multiuser {\it power region} for the fading MAC is defined as
the constitution of all user power-tuples under which a given set of
rates is achievable for all the MTs. Let
$\mv{R}=(R_1,\ldots,R_K)\in\mathbb{R}_{+}^{K}$ denote the vector of
rate requirements for the MTs. The power region is then defined as
\begin{eqnarray} \label{eq:power def SDMA}
\mathcal{P}_{\rm SDMA}\left(\mv{R}\right) \triangleq
\left\{\mv{p}\in \mathbb{R}_{+}^K: \exists \ \{\mv{S}_k\}, {\rm such
\ that} \ \mv{R}\in
\mathcal{C}_{\mathtt{SDMA}}\left(\{\mv{S}_k\}\right),
p_k=\mathtt{Tr}(\mv{S}_k), k=1,\ldots,K \right\},
\end{eqnarray}
for SDMA, and
\begin{eqnarray} \label{eq:power def TDMA}
\mathcal{P}_{\rm TDMA}\left(\mv{R}\right) \triangleq
\left\{\mv{p}\in \mathbb{R}_{+}^K: \exists \ \{\tau_k\},
\{\mv{S}_k\}, {\rm such \ that} \ \mv{R}\in
\mathcal{C}_{\mathtt{TDMA}}\left(\{\tau_k\},\{\mv{S}_k\}\right),
p_k=\mathtt{Tr}(\mv{S}_k), k=1,\ldots,K \right\},
\end{eqnarray}
for TDMA. It is not hard to show that the power regions for both
SDMA and TDMA are convex sets. The power region is illustrated in
Fig. \ref{fig:power region defined} for a two-user fading MAC under
either SDMA or TDMA. The solid line in Fig. \ref{fig:power region
defined} represents the boundary points of the power region, which
correspond to all {\it pareto} optimal power-tuples each minimizing
a weighted-sum of the powers among the MTs. Each boundary point,
e.g., point A as indicated in Fig. \ref{fig:power region defined},
might be characterized by two alternative means described as
follows.

First, because of the convexity of the power region, each boundary
point can be expressed as the solution to a {\it weighted sum-power
minimization} (W-SPmin) problem stated below, for some nonnegative
user weights, $\lambda_k$, $k=1,\ldots,K$. For SDMA, the W-SPmin
problem can be expressed as
\begin{problem} \label{prob:W-SPmin SDMA}
\begin{eqnarray}
\mathop{\mathtt{Minimize}}_{\left\{r_k\right\},\left\{\mv{S}_k\right\}}
& &  \sum_{k=1}^K \lambda_k \mathtt{Tr}\left(\mv{S}_k\right)
\label{eq:weighted power min SDMA}
\\ \mathtt {Subject \ to} & & r_k \geq R_k \ \ \forall k
\label{eq:VR constraint SDMA} \\ & & \mv{r} \in
\mathcal{C}_{\mathtt{SDMA}}(\{\mv{S}_k\}) \label{eq:domain def_1
SDMA}
\\ & & \mv{S}_k \succeq 0 \ \ \ \forall k. \label{eq:domain def_2 SDMA}
\end{eqnarray}
\end{problem}
Notice that $\{r_k\}$ are auxiliary variables. For TDMA, this
W-SPmin problem is given by
\begin{problem} \label{prob:W-SPmin TDMA}
\begin{eqnarray}
\mathop{\mathtt{Minimize}}_{\left\{r_k\right\},\left\{\mv{S}_k\right\},\{\tau_k\}}
& & \sum_{k=1}^K \lambda_k \mathtt{Tr}\left(\mv{S}_k\right)
\label{eq:weighted power min TDMA}
\\ \mathtt {Subject \ to} & & r_k \geq R_k \ \ \forall k
\label{eq:VR constraint TDMA} \\ & & \mv{r} \in
\mathcal{C}_{\mathtt{TDMA}}(\{\tau_k\},\{\mv{S}_k\})
\label{eq:domain def_1 TDMA}
\\ & & \tau_k\geq 0 \ \ \forall k \label{eq:domain def_2 TDMA} \\
& & \sum_{k=1}^K\tau_k=1 \label{eq:domain def_3 TDMA} \\ & &
\mv{S}_k \succeq 0 \ \ \forall k.\label{eq:domain def_4 TDMA}
\end{eqnarray}
\end{problem}

\begin{figure}
\psfrag{a}{$p_1$} \psfrag{b}{$p_2$} \psfrag{c}{$(p_1^*,p_2^*)$}
\psfrag{d}{$\mv{\lambda}^T\mv{p}$} \psfrag{e}{$r_1$}
\psfrag{f}{$r_2$} \psfrag{g}{$(R_1,R_2)$}
\psfrag{h}{$\mv{\rho}^T\mv{r}$}
\begin{center}
\scalebox{1.0}{\includegraphics*[41pt,550pt][471pt,794pt]{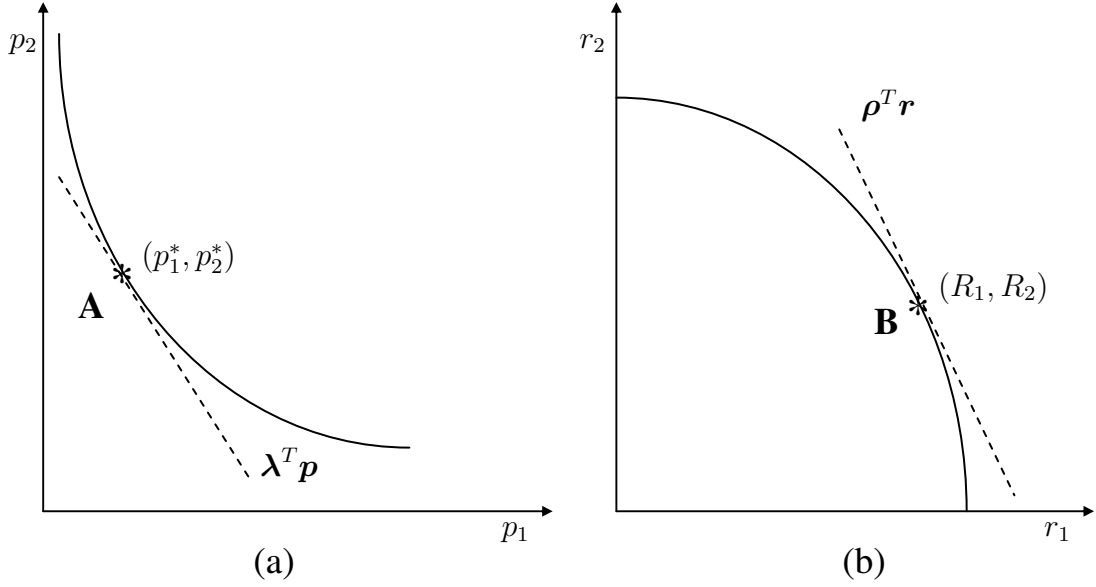}}
\end{center}
\caption{The relationship between each boundary point of the power
region and its corresponding capacity region for a two-user fading
MAC. (a) Power region under the rate constraint $(R_1, R_2)$, where
$(p_1^*, p_2^*)$, represented by point A, is on its boundary and
achieves the minimum weighted sum-power under the weights
$\lambda_1$ and $\lambda_2$, i.e.,
$\lambda_1p_1^*+\lambda_2p_2^*=p^*$; (b) Corresponding capacity
region of $(p_1^*, p_2^*)$ under the weighted sum-power constraint
$\lambda_1p_1+\lambda_2p_2\leq p^*$, where $(R_1, R_2)$, represented
by point B, is on its boundary. Moreover, $(R_1, R_2)$ maximizes the
weighted sum-rate $\rho_1r_1+\rho_2r_2$ for some nonnegative weight
vector $\mv{\rho}$.} \label{fig:power capacity dual}
\end{figure}

By definition of the power region for a given target rate-tuple
$\mv{R}$, each boundary point of the power region, $\mv{p}^*$, under
either SDMA or TDMA, can be expressed as a power-tuple supporting
the target rates in $\mv{R}$ that has the minimum weighted-sum,
$\sum_{k=1}^K \lambda_kp_k^*$, for some weight vector
$\mv{\lambda}$, among all the power-tuples that can support
$\mv{R}$. Alternatively, the connection between the power-tuple
$\mv{p}^*$ on the boundary of the power region and the target rate
$\mv{R}$ can be established by employing the capacity region
concept. Fig. \ref{fig:power capacity dual} gives an illustration
for this important observation. Consider the power region of a
two-user MAC under the rate constraint $(R_1, R_2)$, as shown in
Fig. \ref{fig:power capacity dual} (a). Given $\lambda_1$ and
$\lambda_2$, the solution to the W-SPmin problem is denoted by
$(p_1^*,p_2^*)$, which is represented by point A in Fig.
\ref{fig:power capacity dual} (a), and satisfies
$\lambda_1p_1^*+\lambda_2p_2^*=p^*$, where $p^*$ is the minimum
value of the W-SPmin problem. Thus, the required minimum power-pair
is $(p_1^*,p_2^*)$ for $(R_1, R_2)$. On the other hand, it is not
hard to verify that the rate-pair $(R_1, R_2)$ is on the boundary of
the capacity region for the same MAC under the {\it weighted
sum-power constraint} given by $\lambda_1p_1+\lambda_2p_2\leq p^*$.
This is shown by point B in Fig. \ref{fig:power capacity dual} (b).
Moreover, because of the convexity of the capacity region, $(R_1,
R_2)$ must be the solution to a {\it weighted sum-rate maximization}
(W-SRmax) problem for a given nonnegative user weight vector
$\mv{\rho}$, as shown in Fig. \ref{fig:power capacity dual} (b). The
above observation has an important consequence, i.e., each
power-region boundary point as the solution to the W-SPmin problem
can be equivalently characterized as a boundary point of the
corresponding capacity region under the same weighted sum-power
constraint. As will be shown later in this paper, this result also
motivates the proposed algorithms for the W-SPmin problem under both
SDMA and TDMA.

Alternatively, each boundary point of the power region can also be
considered geometrically as the intersection of a line passing
through the origin (the power-tuple with all zeros) and the boundary
of the power region (see point A in Fig. \ref{fig:power region
defined}). Let each line be characterized as $p_k = \alpha_k P$, for
$k=1, \ldots, K$ and $P \geq 0$. The vector
$\mv{\alpha}=(\alpha_1,\ldots,\alpha_K)\in\mathbb{R}_{+}^K$ is
referred to as the {\it user power-profile} vector in this paper,
and it is assumed that $\sum_{k=1}^{K}\alpha_k=1$. The point where
the line specified by $\mv{\alpha}$ intersects the power region
boundary can be then obtained as the solution to the following
optimization problem, referred to as the {\it sum-power minimization
under the power-profile constraint} (SPmin-PPC). For SDMA, this
problem can be expressed as
\begin{problem} \label{prob:SPMin-PF SDMA}
\begin{eqnarray} \mathop{\mathtt{Minimize}}_{P, \{r_k\}, \left\{\mv{S}_k\right\}} & &  P
\label{eq:power profile min}
\\ \mathtt {Subject \ to} & & r_k \geq R_k \ \
\forall k \label{eq:VR constraint_new}  \\ & & \mv{r} \in
\mathcal{C}_{\mathtt{SDMA}}(\{\mv{S}_k\}) \label{eq:domain def_1
new} \\ & & \mv{S}_k \succeq 0 \ \ \forall k \label{eq:domain
def_2 new} \\ & & \mathtt{Tr}\left(\mv{S}_k\right)\leq \alpha_kP \
\ \forall k \label{eq:power constraint} \\ & & P\geq 0.
\label{eq:power positive constraint}
\end{eqnarray}
\end{problem}
And similarly, the SPmin-PPC problem can be defined for TDMA. The
characterization of the power region via some prescribed
power-profile vector is useful for the BS to regulate the power
consumption of MTs in a desired {\it proportionally-fair} manner.

It is not hard to show that both problems, W-SPmin and SPmin-PPC,
are convex for either SDMA or TDMA, and hence, they can be solved by
applying convex optimization techniques. The following two sections
present the solutions to these problems for SDMA and TDMA,
respectively.

\section{Characterization of Power Region for SDMA} \label{sec:SDMA}

\subsection{Solutions to Weighted Sum-Power Minimization}\label{subsec:weighted sum power}

This subsection considers Problem \ref{prob:W-SPmin SDMA}, the
W-SPmin problem in the case of SDMA. For the special case of a
deterministic (no fading) SISO-MAC, the W-SPmin can be simplified
using the contra-polymatroid structure as proposed in \cite{Tse98a},
\cite{Tse98b}. However, as shown later in this subsection, the
approach in \cite{Tse98a}, \cite{Tse98b} can not be applied here to
handle the more general case of the fading MIMO-MAC. Thus, an
alterative approach is proposed.

{\bf Deterministic SISO-MAC:} Consider a deterministic SISO-MAC
where the channel gain for each MT in (\ref{eq:MIMO MAC}) is a
positive constant, i.e., $|\mv{H}_k(\nu)| = \sqrt{h_k}, \forall
\nu$. In this case, the solutions to the W-SPmin problem are
obtained as follows.


\begin{theorem}\label{theorem:SISO-MAC}
For a deterministic SISO-MAC consisting of $K$ users with channel
gains $h_1,\ldots,h_K$, and rate requirements $R_1,\ldots,R_K$, the
solutions to the W-SPmin problem under SDMA are given by \cite[Lemma
3.2]{Tse98b}:
\begin{eqnarray}\label{eq:optimal powers}
p_{\pi(k)}^*=\left\{
\begin{array}{ll} \frac{\exp\left(2R_{\pi(1)}\right)-1}{h_{\pi(1)}}
& \mathtt{if} \ k=1\\
\frac{\exp\left(2\sum_{i=1}^{k}R_{\pi(i)}\right)-\exp\left(2\sum_{i=1}^{k-1}R_{\pi(i)}\right)}{h_{\pi(k)}}
& k=2,\ldots,K, \end{array} \right.
\end{eqnarray}
where the permutation $\pi$ indicates the optimal decoding order
(user $\pi(1)$ is decoded last and user $\pi(K)$ is decoded first)
at the receiver according to
\begin{eqnarray}\label{eq:decoding orders}
\frac{\lambda_{\pi(1)}}{h_{\pi(1)}}\geq\cdots\geq\frac{\lambda_{\pi(K)}}{h_{\pi(K)}}.
\end{eqnarray}
\end{theorem}

Fig. \ref{fig:power capacity SISO} shows the connection between each
boundary point on the power region and the corresponding capacity
region earlier described in Section \ref{sec:power region} for a
two-user deterministic SISO-MAC. Fig. \ref{fig:power capacity SISO}
(a) shows the power region given rate constraint $(R_1,R_2)$.
Consider a vertex $(p_1^*, p_2^*)$ of this power region, and
arbitrary positive weights $\lambda_1, \lambda_2$ such that
$\frac{\lambda_1}{h_1}>\frac{\lambda_2}{h_2}$. According to Theorem
\ref{theorem:SISO-MAC}, $(p_1^*, p_2^*)$ is the optimal solution to
the W-SPmin problem, and is achievable by decoding order
$2\rightarrow1$ (user 2's message is decoded before user 1's). For
this given choice of $\lambda_1$ and $\lambda_2$, the boundary curve
of the capacity region under the {\it weighted-sum power constraint}
$\lambda_1p_1 + \lambda_2 p_2 \leq p^*$, shown in Fig.
\ref{fig:power capacity SISO} (b), can be represented as the union
of all rate regions, $\mathcal{C}_{\mathtt{SDMA}}(\{p_k\})$ defined
in (\ref{eq:definition C SDMA}), over all values of $p_1$ and $p_2$
that satisfy $\lambda_1p_1+\lambda_2p_2 = p^*$. Each
$\mathcal{C}_{\mathtt{SDMA}}(\{p_k\})$ is a pentagon with two
vertices, each corresponding to one of the two possible decoding
orders among the users \cite{Coverbook}. The fact that user 2's
message must be decoded first to achieve $(R_1, R_2)$ can be
justified by the following observation. It is seen that the decoding
order $2\rightarrow1$ always achieves higher rates for both users
than the other decoding order $1\rightarrow 2$, hence, it must be
the optimal decoding order to achieve the rate target $(R_1,R_2)$ on
the boundary of this capacity region. In general, for a
deterministic SISO-MAC, all rate-tuples on the boundary of the
capacity region under a weighted sum-power constraint can be
achieved by a {\it unique} decoding order for each user. This result
is consistent with Theorem \ref{theorem:SISO-MAC}, i.e., given
$\lambda_k$'s and $h_k$'s, the optimal decoding order of users can
be first resolved by (\ref{eq:decoding orders}), and then their
minimum powers can be found by (\ref{eq:optimal powers}).

\begin{figure}
\begin{center}
\scalebox{1.0}{\includegraphics{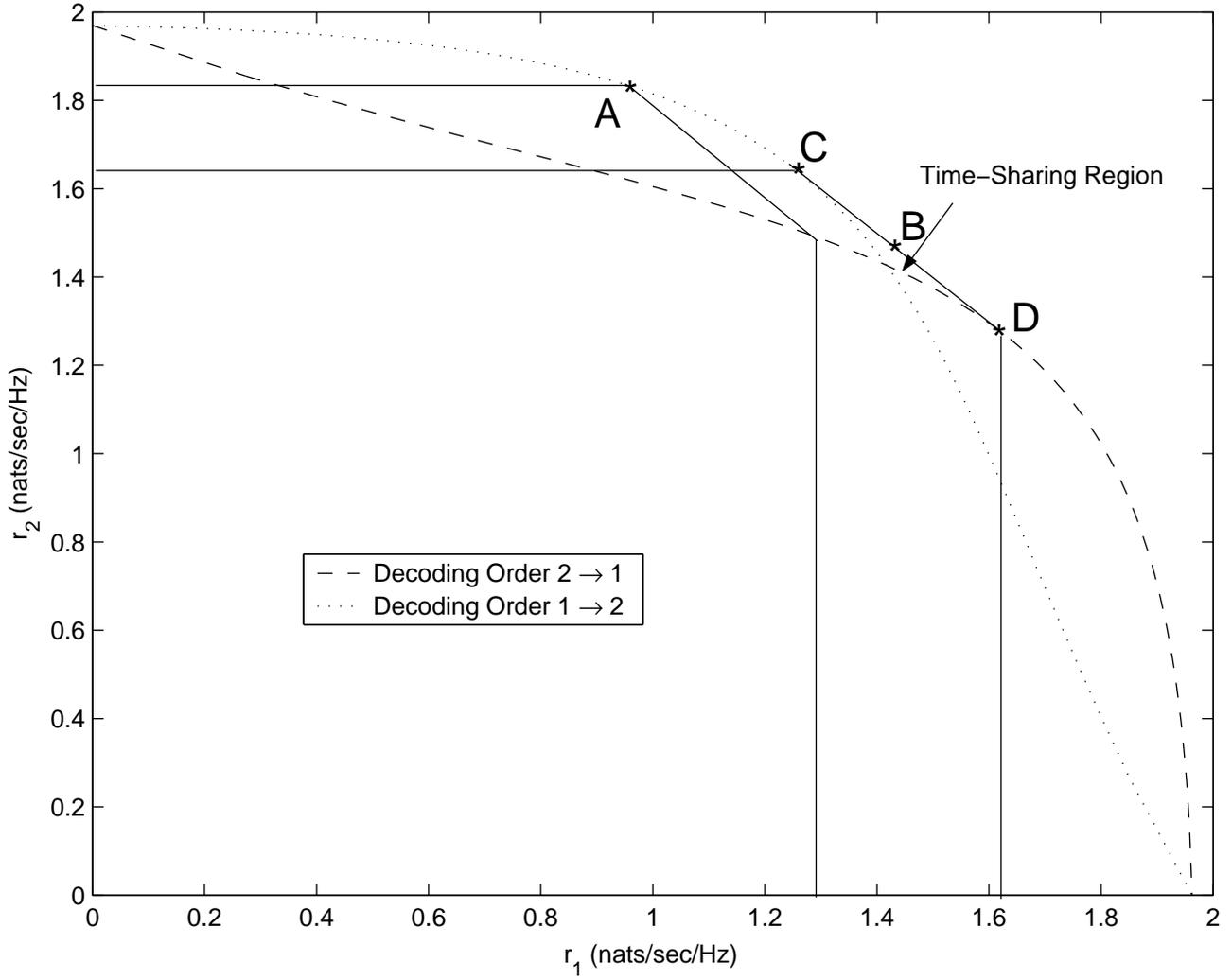}}
\end{center}
\caption{Capacity region for a two-user symmetric-fading SIMO-MAC
under a sum-power constraint: $p_1+p_2\leq 10$. The multiple
access technique is SDMA, and $t_1=t_2=1$, $r=2$.}
\label{fig:capacity region fading SIMO}
\end{figure}

{\bf Fading MIMO-MAC:} Unfortunately, the contra-polymatroid
structure is non-existent for the MAC when there is fading (e.g.,
the fading SISO-MAC) or there are multiple antennas at the receiver
(e.g., the deterministic SIMO-MAC), and hence, it is non-existent
for the general fading MIMO-MAC. Equivalently, the rate-tuples on
the boundary of the capacity region under the weighted sum-power
constraint for a fading MIMO-MAC, unlike the case of a deterministic
SISO-MAC, might not correspond to a unique decoding order for each
user. Fig. \ref{fig:capacity region fading SIMO} illustrates this
fact by showing the capacity region of a two-user fading SIMO-MAC
($t_1=t_2=1, r=2$) under a sum-power constraint $p_1+p_2\leq 10$
(i.e., $\lambda_1=\lambda_2=1$). In this case, the channels
$\{\mv{H}_1(\nu)\}$ and $\{\mv{H}_2(\nu)\}$ are assumed to be
independent vectors each having independent entries distributed as
$\mathcal{CN}(0,1)$. The capacity region for this case is shown to
be symmetric. The dashed line and the dotted line show how two
vertices of the constituting $\mathcal{C}_{\mathtt{SDMA}}(\{p_k\})$
sweep on the boundary of the capacity region as $p_1$ and $p_2$ vary
while their sum is kept equal to 10. It is observed that the
boundary rate-tuples of this capacity region indeed correspond to
different decoding orders; e.g., point A corresponds to the decoding
order $1\rightarrow 2$, while point D does for $2\rightarrow 1$.
There is also a part of the capacity region (e.g., point B is in
this region) that does not consist of any vertices. This part of the
region is referred to as the {\it time-sharing} region and consists
of the -45 degree boundary lines of the constituting
$\mathcal{C}_{\mathtt{SDMA}}(\{p_k\})$. Hence, any point in the
time-sharing region is not achievable by successive decoding given
any fixed decoding orders, and time-sharing the transmission rates
and the decoding orders among two users is required. As a result,
unlike the deterministic SISO-MAC, given $\lambda_k$'s and
$\mv{H}_k(\nu)$'s, the optimal decoding order for the W-SPmin
problem can not be resolved directly for this channel.

One heuristic approach (e.g., \cite{Boche02}-\cite{Jisung05}) for
solving W-SPmin problem for the fading MIMO-MAC under SDMA might be
searching through all possible $K!$ decoding orders and then finding
the optimal decoding order that gives the minimum weighted sum-power
to support the target rates. This approach might work for some
special cases, but as explained below, is problematic in general.

For any fixed decoding order $\pi$, Problem \ref{prob:W-SPmin SDMA}
can be written as
\begin{eqnarray}
\mathop{\mathtt{Minimize}}_{\left\{\mv{S}_k\right\}} & &
\sum_{k=1}^K \lambda_k \mathtt{Tr}\left(\mv{S}_k\right)
\\ \mathtt {Subject \ to} & &  \mathbb{E}_{\nu}\left[ \frac{1}{2}
\log\frac{\left|
 \sum_{i=1}^{k}\mv{H}_{\pi(i)}(\nu)
\mv{S}_{\pi(i)}\mv{H}_{\pi(i)}^{\dagger}(\nu) +\mv{I}
\right|}{\left|\sum_{i=1}^{k-1}\mv{H}_{\pi(i)}(\nu)\mv{S}_{\pi(i)}\mv{H}_{\pi(i)}^{\dagger}(\nu)
+\mv{I}\right|} \right]\geq R_{\pi(k)} \ \ \forall k \label{eq:rate
given decoding order}
\\ & & \mv{S}_k \succeq 0 \ \ \ \forall k,
\end{eqnarray}
where the left-hand-side (LHS) of (\ref{eq:rate given decoding
order}) is the achievable rate for MT $\pi(k)$, denoted as
$r^{(\pi)}_{\pi(k)}$, under the decoding order $\pi$. Except for MT
$\pi(1)$, the rate constraints in (\ref{eq:rate given decoding
order}) are not convex, rendering the above optimization problem
non-convex in general and, hence, it can not be solved efficiently.
A suboptimal method that approximately solves this problem is
described as follows. Starting from the last decoded MT $\pi(1)$,
the method minimizes the power required to maintain the target rate
for that MT, while considering MTs that have yet been decoded as
interference. For example, for the two-user case and the decoding
order of MT 2 followed by MT 1, the method first determines
$\mv{S}_1$ with the minimum $p_1$ that satisfies
$\mathbb{E}_{\nu}[\frac{1}{2} \log
|\mv{H}_1(\nu)\mv{S}_1\mv{H}_1^{\dagger}(\nu)+\mv{I}|]\geq R_1$ and
then fixes $\mv{S}_1$ and determines $\mv{S}_2$ with the minimum
power $p_2$ that satisfies $\mathbb{E}_{\nu}[\frac{1}{2} \log
|\mv{H}_1(\nu)\mv{S}_1\mv{H}_1^{\dagger}(\nu)+\mv{H}_2(\nu)\mv{S}_2\mv{H}_2^{\dagger}(\nu)
+ \mv{I}|] \geq R_1+R_2$. Each of these two optimizations are convex
and, hence, they both can be solved efficiently. The above algorithm
is referred to as the {\it greedy} algorithm since each MT simply
minimizes its own transmit power. For the special case of the fading
SISO-MAC and SIMO-MAC where each $\mv{S}_k$ is simply a scalar and
is equal to $p_k$, the obtained $p_1$ and $p_2$ via the greedy
algorithm are in fact optimal and minimize the weighted sum-power
for any weights $\mv{\lambda}$ under the given decoding order. This
is because from (\ref{eq:rate given decoding order}) it can be shown
that the minimum $p_{\pi(k)}$ required to support $R_{\pi(k)}$ for
user $\pi(k)$ is an increasing function of the powers for the
not-yet-decoded users, $p_{\pi(1)},\ldots,p_{\pi(k-1)}$. However,
for the general fading MIMO-MAC, the above greedy algorithm might
not be optimal because each MT now can adjust its covariance matrix
to balance between minimizing its own transmit power and reducing
the interference it causes to the users decoded earlier in the
order.

Nevertheless, even if the W-SPmin problem can be solved for each
decoding order, the obtained powers that have the minimum
weighted-sum among all decoding orders might still be suboptimal.
This can occur when the target rate-tuple does not correspond to a
unique optimal decoding order, e.g., the rate-pair B in Fig.
\ref{fig:capacity region fading SIMO} that is on the boundary of the
time-sharing region.

From the above discussions, it follows that for the fading MIMO-MAC
in general, the decoding order and the transmit covariance matrices
for the MTs need to be jointly optimized for solving the W-SPmin
problem under SDMA. This motivates the algorithm presented next.

{\bf Proposed Algorithm:} The proposed algorithm for Problem
\ref{prob:W-SPmin SDMA} is based on its Lagrangian \cite{Boydbook},
which is given below:
\begin{eqnarray}\label{eq:Lagrange power min SDMA}
\mathcal{L}(\{\mv{S}_k\},\{r_k\},\mv{\mu})=
\sum_{k=1}^{K}\lambda_k \mathtt{Tr}(\mv{S}_k)-\sum_{k=1}^{K}\mu_k
(r_k-R_k),
\end{eqnarray}
where $\mv{\mu}=(\mu_1,\ldots,\mu_K)\in\mathbb{R}_{+}^K$ denotes
the vector of dual variables associated with the rate inequality
constraints in (\ref{eq:VR constraint SDMA}). The variables,
$\{\mv{S}_k\}$ and $\{r_k\}$ belong to the set denoted by
$\mathcal{D}$, which is specified by the remaining constrains in
(\ref{eq:domain def_1 SDMA}) and (\ref{eq:domain def_2 SDMA}).
Then the Lagrange dual function \cite{Boydbook} is defined as
\begin{eqnarray}\label{eq:Lagrange dual power min SDMA}
g(\mv{\mu})=\min_{\left\{\mv{S}_k, r_k\right\}\in \mathcal{D}}
\mathcal{L}(\{\mv{S}_k\},\{r_k\},\mv{\mu}).
\end{eqnarray}
The dual function serves as a lower bound on the optimal value of
the original (primal) problem, denoted by $q^{*}$, i.e., $q^*\geq
g(\mv{\mu}), \forall \mv{\mu}$ \cite{Boydbook}. The dual problem is
then defined as $\max_{\mv{\mu} \succeq 0 } g(\mv{\mu})$
\cite{Boydbook}. Let the optimal value of the dual problem be
denoted by $d^*$ that is achievable by the optimal dual variables
$\mv{\mu}^*$, i.e., $d^*= g(\mv{\mu}^{*})$. For a convex
optimization problem, the Slater's condition states that the duality
gap, $q^* - d^* \geq 0$, is indeed zero if the primal problem has a
feasible solution set \cite{Boydbook}. By using sufficiently large
user powers, the set $\mathcal{C}_{\mathtt{SDMA}}(\{\mv{S}_k\})$ in
(\ref{eq:domain def_1 SDMA}) can be made arbitrarily large to
contain any finite rate target $\mv{R}$ as an interior point. In
other words, we can always find a feasible set $\{\mv{S}_k\}$ that
satisfies any given rate constraint $\mv{R}$ for Problem
\ref{prob:W-SPmin SDMA}. Thus, the Slater's condition holds and the
duality gap is zero for Problem \ref{prob:W-SPmin SDMA}. This result
suggests that $q^*$ can be obtained by first minimizing the
Lagrangian $\mathcal{L}$ to obtain the dual function $g(\mv{\mu})$
for some given $\mv{\mu}$, and then maximizing $g(\mv{\mu})$ over
all possible values $\mv{\mu}$.

First, consider the minimization of $\mathcal{L}$ to obtain the
dual function $g(\mv{\mu})$. In this case, $\mv{\mu}$ is fixed and
the variables are $\{\mv{S}_k\}$ and $\{r_k\}$. From
(\ref{eq:Lagrange power min SDMA}), it follows that the
minimization of $\mathcal{L}$ can be rewritten as the following
equivalent problem:
\begin{eqnarray}
\mathop{\mathtt{Minimize}}_{\left\{r_k\right\},\left\{\mv{S}_k\right\}}
& & \sum_{k=1}^{K}\lambda_k
\mathtt{Tr}(\mv{S}_k)-\sum_{k=1}^{K}\mu_k r_k \label{eq:general
minpower MAC}
\\ \mathtt {Subject \ to} & & \mv{r} \in
\mathcal{C}_{\mathtt{SDMA}}(\{\mv{S}_k\}) \label{eq:polymatroid
constraint}
\\ & & \mv{S}_k \succeq 0 \ \ \forall k.
\end{eqnarray}
From the definition of $\mathcal{C}_{\mathtt{SDMA}}$ in
(\ref{eq:definition C SDMA}), there are $2^K-1$ rate constraints
implied by (\ref{eq:polymatroid constraint}), which are difficult to
be incorporated directly into the optimization. In order to simplify
the problem, the following theorem is utilized to remove these
constraints in (\ref{eq:polymatroid constraint}):
\begin{theorem}\label{theorem:polymatroid}
For any permutation $\pi$ over $\{1,\ldots,K\}$ and a fixed set of
covariance matrices $\{\mv{S}_k\}$, $\mv{r}^{(\pi)}$ defined as
\begin{eqnarray}
r^{(\pi)}_{\pi(k)} = \mathbb{E}_{\nu}\left[ \frac{1}{2}
\log\frac{\left|
 \sum_{i=1}^{k}\mv{H}_{\pi(i)}(\nu)
\mv{S}_{\pi(i)}\mv{H}_{\pi(i)}^{\dagger}(\nu) +\mv{I}
\right|}{\left|\sum_{i=1}^{k-1}\mv{H}_{\pi(i)}(\nu)\mv{S}_{\pi(i)}\mv{H}_{\pi(i)}^{\dagger}(\nu)
+\mv{I}\right|} \right]
\end{eqnarray}
is a {\it vertex} of the polymatroid
$\mathcal{C}_{\mathtt{SDMA}}(\{\mv{S}_k\})$ in $\mathbb{R}^K_+$.
Furthermore, for any $\mv{\rho} \succeq 0$, the solution to the
following W-SRmax problem:
\begin{eqnarray}
\mathop{\mathtt{Maximize}}_{\{r_k\}} & & \sum_{k=1}^K \rho_k r_k
\\ \mathtt {Subject \ to} & & \mv{r} \in
\mathcal{C}_{\mathtt{SDMA}}(\{\mv{S}_k\}),
\end{eqnarray}
is attained by a vertex $\mv{r}^{(\pi^*)}$, where $\pi^*$ is such
that $\rho_{\pi^*(1)} \geq \rho_{\pi^*(2)} \geq \ldots \geq
\rho_{\pi^*(K)}$.
\end{theorem}
\begin{proof}
Please refer to \cite[Lemma 3.10]{Tse98a}.
\end{proof}

Notice that Theorem \ref{theorem:polymatroid} holds for any given
$\left\{\mv{S}_k\right\}$. Furthermore, since minimization of
$-\sum_k\mu_k r_k$ is equivalent to maximization of $\sum_k\mu_k
r_k$, using Theorem \ref{theorem:polymatroid}, the rate constraints
in (\ref{eq:polymatroid constraint}) can be safely removed and the
problem in (\ref{eq:general minpower MAC}) can be simplified as
\begin{eqnarray}
\mathop{\mathtt{Minimize}}_{\left\{\mv{S}_k\right\}} & &
\sum_{k=1}^{K}\lambda_k
\mathtt{Tr}(\mv{S}_k)-\sum_{k=1}^{K}\mu_{\pi(k)}\mathbb{E}_{\nu}\left[
\frac{1}{2} \log\frac{\left|
 \sum_{i=1}^{k}\mv{H}_{\pi(i)}(\nu)
\mv{S}_{\pi(i)}\mv{H}_{\pi(i)}^{\dagger}(\nu) +\mv{I}
\right|}{\left|\sum_{i=1}^{k-1}\mv{H}_{\pi(i)}(\nu)\mv{S}_{\pi(i)}\mv{H}_{\pi(i)}^{\dagger}(\nu)
+\mv{I}\right|} \right] \label{eq:mac polymatroid new}
\\ \mathtt {Subject \ to} & & \mv{S}_k \succeq 0 \ \ \forall k,
\end{eqnarray}
where $\pi$ is a permutation such that $\mu_{\pi(1)} \geq \cdots
\geq \mu_{\pi(K)}$. By rearranging the terms regarding user rates in
(\ref{eq:mac polymatroid new}), the above problem becomes the
minimization of
\begin{eqnarray}\label{eq:mac polymatroid}
\sum_{k=1}^K \lambda_k
\mathtt{Tr}(\mv{S}_k)-\sum_{k=1}^{K}\left(\mu_{\pi(k)} -
\mu_{\pi(k+1)}\right)\mathbb{E}_{\nu}\left[\frac{1}{2} \log
\left|\sum_{i=1}^{k} \left(\mv{H}_{\pi(i)}(\nu)
\mv{S}_{\pi(i)}\mv{H}_{\pi(i)}^{\dagger}(\nu)\right) +\mv{I}
\right| \right],
\end{eqnarray}
with only optimization variables $ \mv{S}_k \succeq 0$, $\forall k$,
and  $\mu_{\pi(K+1)}\triangleq 0$. Since the above problem is convex
with a twice differentiable objective function and positive
semi-definite constraints, it can be solved numerically, e.g., by
the interior-point method \cite{Boydbook}.

Next, the dual function $g(\mv{\mu})$ is maximized over all possible
values $\mv{\mu}$. The search of $\mv{\mu}$ towards its optimal
value $\mv{\mu}^*$  can be done, e.g., by the ellipsoid method
\cite{BGT81}, which utilizes the fact that the vector $\mv{\theta}$,
defined as $\theta_k = R_k - r'_k$ for $k=1,\ldots,K$, is a {\it
sub-gradient} of $g(\mv{\mu})$ at any $\mv{\mu}$, where
$\{\mv{S}'_k$\} and $\{r'_k\}$ are the minimizers of
$\mathcal{L}(\{\mv{S}_k\}, \{r_k\}, \mv{\mu})$ obtained via solving
(\ref{eq:mac polymatroid}), i.e., $\mathcal{L}(\{\mv{S}'_k\},
\{r'_k\}, \mv{\mu}) = g(\mv{\mu})$.

%

\begin{remark}
It is noted that the algorithm proposed in \cite[Algorithm
5.3]{Tse98a} can also be modified to determine $\mv{\mu}^{*}$ for
the problem at hand. However, from programming implementations, it
is observed that this method may exhibit oscillation when some of
$\mu_k^*$'s happen to be equal. In contrast, the ellipsoid method is
more suitable because of its robust and superior convergence
behavior.
\end{remark}

The complete algorithm for Problem \ref{prob:W-SPmin SDMA} in the
case of SDMA is summarized below.
\begin{algorithm}\label{algorithm:minpower MAC SDMA}
\
\
\begin{itemize}
\item {\bf Given} an ellipsoid $\mathcal{E}[0]\subseteq
\mathbb{R}^K$, centered at $\mv{\mu}[0]$ and containing the
optimal dual solution $\mv{\mu}^{*}$.
\item {\bf Set} $i=0$.
\item {\bf Repeat}
\begin{itemize}
\item [1.] For given $\mv{\mu}[i]$, solve the problem given in (\ref{eq:mac polymatroid}) to obtain an optimal solution set
$\{\mv{S}_k[i]\}$ and $\{r_k[i]\}$ that minimizes
$\mathcal{L}(\{\mv{S}_k\}, \{r_k\}, \mv{\mu}[i])$ over
$\mathcal{D}$;
\item[2.] Update
the ellipsoid $\mathcal{E}[i+1]$ based on $\mathcal{E}[i]$ and the
sub-gradients $\theta_k[i] = R_k - r_k[i], k=1,\ldots,K$. Set
$\mv{\mu}[i+1]$ as the center of the new ellipsoid
$\mathcal{E}[i+1]$;\footnote{Notice that when locating the center of
a new ellipsoid, we need to add the constraint that $\mv{\mu}\succeq
0$.}
\item[3.] Set $i\leftarrow i+1$.
\end{itemize}
\item {\bf Until} the stopping criteria for the ellipsoid method
is met.
\end{itemize}
\end{algorithm}
One possible method to obtain the initial ellipsoid $\mathcal{E}[0]$
that contains the optimal dual solution $\mv{\mu}^{*}$ is given in
Appendix \ref{appendix:initial ellipsoid}.

The primal-dual approach used for solving Problem \ref{prob:W-SPmin
SDMA} can be explained by the connection between the power region
and the capacity region as described in Section \ref{sec:power
region}. In Fig. \ref{fig:power capacity dual}, it is observed that
each power-tuple on the boundary of the power region for a given
target rate-tuple, as the solution to Problem \ref{prob:W-SPmin
SDMA} for a given weight vector $\mv{\lambda}$, defines a capacity
region that contains the target rate-tuple on its boundary.
Consequently, the target rate-tuple can be expressed as the solution
to a W-SRmax problem for some unknown user weight vector $\mv{\rho}$
over this capacity region. Clearly, the primal-dual approach
establishes the above connection by finding the optimal dual
variable $\mv{\mu}^*$ that is simply one candidate for the unknown
weight vector $\mv{\rho}$ in this W-SRmax problem. From (\ref{eq:mac
polymatroid}), it follows that the optimal decoding order of each MT
is given by the magnitude of $\mu_k^*$, i.e., the optimal decoding
order $\pi$ satisfies $\mu_{\pi(1)}^*\geq\cdots\geq \mu_{\pi(K)}^*$.
Hence, the proposed algorithm successfully jointly optimizes the
decoding order and the transmit covariance matrices of the MTs by
exploiting the duality between the power region and the capacity
region.

{\bf Uniqueness of Solutions:} So far, the proposed algorithm
determines the optimal value of the primal problem $q^*$ (equal to
that of the dual problem $d^*$), the corresponding primal variables
$\{\mv{S}_k^*\}$ and $\{r_k^*\}$ (it is yet claimed that these
primal variables are the primal optimal solutions), and the dual
optimal solutions $\mv{\mu}^{*}$ that satisfy,
\begin{eqnarray}\label{eq:optimal value SDMA}
q^*= d^* = \sum_{k=1}^{K}\lambda_k
\mathtt{Tr}(\mv{S}_k^*)-\sum_{k=1}^{K}\mu_k^* (r_k^*-R_k).
\end{eqnarray}
In the following, the issue on the uniqueness of these solutions is
addressed. While uniqueness of the dual optimal variables
$\mv{\mu}^{*}$ is not an issue in the convergence of the proposed
algorithm,\footnote{By the primal-dual approach and the connection
between the power region and the capacity region, it follows that
$\mv{\mu}^*$ can be viewed as the weight vector $\mv{\rho}$ that
attains the given rate requirements $\mv{R}$ as the solution to the
W-SRmax problem over the corresponding capacity region. For the
fading MIMO-MAC, as shown in Fig. \ref{fig:capacity region fading
SIMO}, in general there is no ``sharp'' vertex with multiple tangent
lines on the boundary of the capacity region under a weighted
sum-power constraint. As a result, it can be inferred that
$\mv{\mu}^{*}$ is unique for any rate-tuple on the boundary and,
hence, the uniqueness of $\mv{\mu}^*$ is in general ensured.}
uniqueness of the primal variables, $\{\mv{S}_k^*\}$ and
$\{r_k^*\}$, plays an important role in obtaining valid solutions
for the W-SPmin problem. Since a primal-dual approach is used, the
obtained primal variables that minimize the Lagrangian at
$\mv{\mu}^*$ might not satisfy the rate constraints in (\ref{eq:VR
constraint SDMA}). Notice that these variables are minimizers of the
Lagrangian and are not necessarily the primal optimal solutions.
However, according to Karush-Kuhn-Tucker (KKT) optimality conditions
\cite{Boydbook}, the primal optimal solutions also minimize the
Lagrangian at $\mv{\mu}^*$. Hence, if these Lagrangian minimizers
can be proven to be unique, it follows that they satisfy the rate
constraints in (\ref{eq:VR constraint SDMA}) automatically.

\begin{theorem}\label{theorem:uniqueness of Sk}
The primal optimal solutions $\{\mv{S}_k^*\}$ for Problem
\ref{prob:W-SPmin SDMA} under SDMA is unique.
\end{theorem}
\begin{proof}
Please refer to Appendix \ref{appendix:proof Sk}.
\end{proof}

If all $\mu_k^*$'s are positive and distinct, $\mv{r}^*$ (e.g.,
shown by Point A in Fig. \ref{fig:capacity region fading SIMO}) that
maximizes $\sum_k \mu^*_k r_k$ over $\mathcal{C}_{\mathtt{SDMA}}(
\{\mv{S}^*_k\})$ will be one unique vertex of
$\mathcal{C}_{\mathtt{SDMA}}( \{\mv{S}^*_k\})$, which itself is also
unique according to Theorem \ref{theorem:uniqueness of Sk}.  In this
case, from the KKT conditions, $r_k^*$'s automatically satisfy the
rate constraints in (\ref{eq:VR constraint SDMA}). However, if
$\mu_k^*$'s in some subset $\mathcal{J}\subseteq\{1,\ldots,K\}$ are
positive and equal, $r_k^*$'s for the users in the set $\mathcal{J}$
may not be unique and consequently they may not satisfy the rate
constraints in (\ref{eq:VR constraint SDMA}). This can be shown by
point B, C and D in Fig. \ref{fig:capacity region fading SIMO} where
$\mu_1^*=\mu_2^*$.  Any point on the straight line between point C
and D maximizes $\sum_k \mu^*_k r_k$ over
$\mathcal{C}_{\mathtt{SDMA}}(\{\mv{S}^*_k\})$ because all these
rate-pairs have the same sum-rate. Hence, if point B is the rate
demand for our problem, because the simplified optimization in
(\ref{eq:mac polymatroid}) always tends to use a vertex of the
capacity region as the solution for $\mv{r}^*$, the proposed
algorithm would converge to either point C or D as $\mv{r}^*$, which
clearly does not satisfy the rate constraints. However, this is not
an issue in the convergence of the proposed algorithm. As far as
$\{\mv{S}_k^*\}$ is unique, the target rate-pair is ensued to be
some convex combination of at most $K$ vertices of the unique
$\mathcal{C}_{\mathtt{SDMA}}(\{\mv{S}_k^*\})$.

\subsection{Solutions to Sum-Power Minimization Under Power-Profile Constraint}\label{subsec:power profile}
This subsection presents the solutions to Problem \ref{prob:SPMin-PF
SDMA}, the SPmin-PPC problem under SDMA, where the transmit power of
each MT is regulated according to a given power-profile vector
$\mv{\alpha}$. The proposed algorithm is also based on a Lagrange
primal-dual approach. The Lagrangian of the primal problem in
(\ref{eq:power profile min}) can be written as
\begin{eqnarray}\label{eq:Lagrange power
min_new} \mathcal{L}(P, \{\mv{S}_k\},\{r_k\},\mv{\mu},\mv{\delta}) =
P+\sum_{k=1}^{K}\delta_k( \mathtt{Tr}(\mv{S}_k)-\alpha_kP)
-\sum_{k=1}^{K}\mu_k (r_k-R_k),
\end{eqnarray}
where $\mv{\mu}=(\mu_1,\ldots,\mu_K)\in\mathbb{R}_{+}^K$ and
$\mv{\delta}=(\delta_1,\ldots,\delta_K)\in\mathbb{R}^K_+$ are dual
variables associated with the inequality constraints in (\ref{eq:VR
constraint_new}) and (\ref{eq:power constraint}), respectively. Let
$\mathcal{F}$ denote the set of primal variables specified by the
remaining constraints in (\ref{eq:domain def_1 new}),
(\ref{eq:domain def_2 new}) and (\ref{eq:power positive
constraint}), the Lagrange dual function can be then expressed as
\begin{eqnarray}\label{eq:Lagrange dual power min new}
g(\mv{\mu},\mv{\delta})=\min_{\left\{P, \mv{S}_k,r_k \right
\}\in\mathcal{F}} P \left(1-\sum_{k=1}^{K}\delta_k\alpha_k\right) +
\sum_{k=1}^{K}\delta_k \mathtt{Tr}(\mv{S}_k) -\sum_{k=1}^{K}\mu_k
(r_k-R_k).
\end{eqnarray}
From (\ref{eq:Lagrange dual power min new}), it is necessary that
$1-\sum_{k=1}^{K}\delta_k\alpha_k\geq0$ for the dual function to be
bounded from below. In the case of
$1-\sum_{k=1}^{K}\delta_k\alpha_k=0$, the optimal $P$ that minimizes
the Lagrangian over $\mathcal{F}$ can take any positive value; while
in the case of $1-\sum_{k=1}^{K}\delta_k\alpha_k>0$, the optimal $P$
must be equal to zero. Thus, in both case, the term associated with
$P$ in (\ref{eq:Lagrange dual power min new}) is indeed zero and,
hence, can be removed from this point onward. The optimal value of
$P$ can be then obtained as
\begin{eqnarray}
P^*=\max_{\mv{\mu} \succeq 0, \mv{\delta}\succeq 0,
\sum_{k=1}^{K}\delta_k\alpha_k \leq 1} g(\mv{\mu},\mv{\delta}).
\end{eqnarray}
Similar algorithm like Algorithm \ref{algorithm:minpower MAC SDMA}
for Problem \ref{prob:W-SPmin SDMA} can be readily developed for
solving this problem. It can be shown that for the problem at hand,
$g(\mv{\mu},\mv{\delta})$ has sub-gradients, $\mv{\theta}$ and
$\mv{\zeta}$ for $\mv{\mu}$ and $\mv{\delta}$, respectively, which
are defined as $\theta_k=R_k-r'_k$ and $\zeta_k =
\mathtt{Tr}(\mv{S}'_{k})$, $k=1,\ldots,K$, where  $\{\mv{S}'_k\}$
and $\{r'_k\}$ are the Lagrangian minimizers that satisfy
$\mathcal{L}(\{\mv{S}'_k\}, \{r'_k\}, \mv{\mu}, \mv{\delta}) =
g(\mv{\mu},\mv{\delta})$.

The convergence of the algorithm for this problem and the uniqueness
of the solutions are similar as Algorithm \ref{algorithm:minpower
MAC SDMA}. Let $\left\{\mv{S}_k^*\right\}$ and $\mv{\delta}^*$
denote the corresponding primal and dual optimal solutions. It is
worth mentioning here that the obtained primal solution for this
problem $P^*$ might not be equal to the user sum-power
$\sum_{k=1}^{K} \mathtt{Tr}(\mv{S}_k^*)$ since some of the power
constraints in (\ref{eq:power constraint}), $\mathtt{Tr}(\mv{S}_k)
\leq \alpha_k P$, may not be active in general.\footnote{For
example, in the case of two-user deterministic SISO-MAC as shown in
Fig. \ref{fig:power capacity SISO} (a), if the given power-profile
vector $\mv{\alpha}$ for Problem \ref{prob:SPMin-PF SDMA} is such
that the intersected boundary power-tuple
$(\alpha_1P^*,\alpha_2P^*)$ of the power region
 is located on the vertical (or
horizontal) boundary segment of the power region, the solutions to
Problem \ref{prob:SPMin-PF SDMA} will converge to the upper (or
lower) vertex $(p_1^*,p_2^*)$ of the power region for which,
clearly, $p_1^*=\alpha_1P^*$, but $p_2^*<\alpha_2P^*$. Thus,
$P^*\neq p_1^*+p_2^*$.} Actually,
$P^*=\sum_{k=1}^{K}\delta_k^*\mathtt{Tr}(\mv{S}_k^*)$ shown as
follows. Since $P^*
>0$,  it can be verified from (\ref{eq:Lagrange dual power min new})
that $\sum_{k=1}^{K}\delta^*_k\alpha_k =1$. Moreover, by the KKT
conditions for the power constraints,
$\sum_{k=1}^{K}\delta^*_k(\mathtt{Tr}(\mv{S}^*_k)-\alpha_kP^*)=0$.
Hence,
$\sum_{k=1}^{K}\delta^*_k\mathtt{Tr}(\mv{S}^*_k)=\sum_{k=1}^K\delta^*_k\alpha_kP^*=P^*$.

The primal-dual approach used for solving the SPmin-PPC problem is
also based on the connection between the power region and the
corresponding capacity region, similar as that for the W-SPmin
problem. However, their difference lies in that for the SPmin-PPC
problem, with reference to Fig. \ref{fig:power capacity dual}, the
weight vector for characterizing the solution on the boundary of the
power region, $\mv{\lambda}$, and the weight vector for the
corresponding capacity region, $\mv{\rho}$, are both unknown and,
hence, they need to be found under the given power-profile vector
$\mv{\alpha}$ as the corresponding optimal dual solution
$\mv{\delta}^*$ and $\mv{\mu}^*$, respectively. In contrast, for the
W-SPmin problem, only the unknown $\mv{\rho}$ needs to be found as
the optimal dual solution $\mv{\mu}^*$ because $\mv{\lambda}$ is
already given.

\section{Characterization of Power Region for TDMA} \label{sec:TDMA}

This section considers the characterization of the power region
defined in (\ref{eq:power def TDMA}) for the fading MIMO-MAC under
TDMA, and for brevity only the W-SPmin problem (Problem
\ref{prob:W-SPmin TDMA}) is investigated. The alternative means for
characterizing the power region based on the power-profile vector,
i.e., the SPmin-PPC problem under TDMA, is omitted since it can be
readily obtained given the techniques developed in Section
\ref{subsec:power profile} for the case of SDMA. The proposed
algorithm for Problem \ref{prob:W-SPmin TDMA} is also based on the
Lagrange primal-dual approach.

First, it is noted that the constraints in (\ref{eq:VR constraint
TDMA}) and (\ref{eq:domain def_1 TDMA})  in Problem
\ref{prob:W-SPmin TDMA}  can be combined and thus simplified,
given the fact that in (\ref{eq:definition C TDMA}), the
inequality constraints are always satisfied with equalities for
power minimization. Hence, the Lagrangian of the primal problem
can be written as
\begin{eqnarray}\label{eq:Lagrange power min TDMA}
\mathcal{L}(\{\tau_k\},\{\mv{S}_k\},\mv{\mu})=
\sum_{k=1}^{K}\lambda_k\mathtt{Tr}(\mv{S}_k)-\sum_{k=1}^{K}\mu_k
\left(\tau_k\mathbb{E}_{\nu}\left[\frac{1}{2} \log \left|
\mv{H}_k(\nu)\frac{\mv{S}_k}{\tau_k}\mv{H}_k^{\dag}(\nu)+\mv{I}\right|\right]-R_k\right),
\end{eqnarray}
where $\mv{\mu}=(\mu_1,\ldots,\mu_K)\in\mathbb{R}_{+}^K$ denotes the
dual variables associated with the constraints in (\ref{eq:VR
constraint TDMA}). The variables, $\{\mv{S}_k\}$ and $\{\tau_k\}$,
belong to the set denoted by $\mathcal{G}$, which is specified by
the remaining constrains in (\ref{eq:domain def_2 TDMA}),
(\ref{eq:domain def_3 TDMA}) and (\ref{eq:domain def_4 TDMA}). Also
note that the Lagrangian is a convex function of both $\{\mv{S}_k\}$
and $\{\tau_k\}$. By changing the variables as
$\frac{\mv{S}_k}{\tau_k} \longmapsto \mv{W}_k, k=1,\ldots,K$, the
Lagrangian can be rewritten as
\begin{eqnarray}\label{eq:Lagrange power min TDMA rewritten}
\mathcal{L}(\{\tau_k\},\{\mv{W}_k\},\mv{\mu})=
\sum_{k=1}^{K}\lambda_k\tau_k\mathtt{Tr}(\mv{W}_k)-\sum_{k=1}^{K}\mu_k
\left(\tau_k\mathbb{E}_{\nu}\left[\frac{1}{2} \log \left|
\mv{H}_k(\nu)\mv{W}_k\mv{H}_k^{\dag}(\nu)+\mv{I}\right|\right]-R_k\right).
\end{eqnarray}

For $k=1,\ldots,K$, define
\begin{eqnarray}\label{eq:zk}
z_k(\mv{W}_k,\mu_k)\triangleq \lambda_k
\mathtt{Tr}(\mv{W}_k)-\mu_k \mathbb{E}_{\nu}\left[\frac{1}{2} \log
\left|
\mv{H}_k(\nu)\mv{W}_k\mv{H}_k^{\dag}(\nu)+\mv{I}\right|\right],
\end{eqnarray}
and
\begin{eqnarray}\label{eq:zk hat}
\hat{z}_k(\mu_k)=\min_{\mv{W}_k\succeq 0} z_k(\mv{W}_k,\mu_k).
\end{eqnarray}
From (\ref{eq:Lagrange power min TDMA rewritten}), (\ref{eq:zk}),
and (\ref{eq:zk hat}), the Lagrange dual function can be expressed
as
\begin{eqnarray}\label{eq:Lagrange dual power min TDMA}
g(\mv{\mu})&=& \min_{\left\{\mv{W}_k, \tau_k\right\}\in
\mathcal{G}}\mathcal{L}(\{\tau_k\},\{\mv{W}_k\},\mv{\mu}) \\ &=&
\min_{\{\tau_k\}: \tau_k\geq 0 \ \forall k, \sum_{k=1}^K\tau_k=1}
\sum_{k=1}^{K}\tau_k
\hat{z}_k(\mu_k)+\sum_{k=1}^{K}\mu_kR_k.\label{eq:Lagrange dual
power min TDMA}
\end{eqnarray}
The optimal value of the primal problem, denoted by $q^{*}$, can be
then obtained as
\begin{eqnarray}
q^{*}&=&\max_{\mv{\mu} \succeq 0} g(\mv{\mu}) \\ &\triangleq&
\sum_{k=1}^{K}\tau_k^* \hat{z}_k(\mu_k^*)+\sum_{k=1}^{K}\mu_k^*R_k.
\end{eqnarray}

\begin{theorem}\label{theorem:TDMA}
If $\{\mv{S}_k^*\}, \{\tau_k^*\}, \{r_k^*\}$ are the optimal
primal solutions and $\{\mu_k^*\}$ are the optimal dual solutions
for Problem \ref{prob:W-SPmin TDMA} under the strictly positive
weight vector $\mv{\lambda}$ and rate target $\mv{R}$, they must
satisfy
\begin{equation}
\lambda_kp_k^*-\mu_k^*r_k^*=c^*\tau_k^*,  \ \ \ \ k=1,\ldots,K,
\end{equation}
where $p_k^*=\mathtt{Tr}(\mv{S}_k^*)$,
$r_k^*=\tau_k^*\mathbb{E}_{\nu}\left[\frac{1}{2} \log \left|
\mv{H}_k(\nu)\frac{\mv{S}_k^*}{\tau_k^*}\mv{H}_k^{\dag}(\nu)+\mv{I}\right|\right]=R_k$,
and $c^*$ is a constant.
\end{theorem}
\begin{proof}
Since $\mv{R}$ and $\mv{\lambda}$ are both strictly positive, then
so are the obtained solutions $\{\tau_k^*\}$. As a result, it is
necessary to have $\hat{z}_1(\mu_1^*)=\cdots=\hat{z}_K(\mu_K^*)$,
otherwise the minimization in (\ref{eq:Lagrange dual power min
TDMA}) must lead to only one user assigned with the total time slot,
i.e., $\tau_{k'}^*=1$ and $\tau_{k}^*=0, k\neq k'$ where
$k'=\arg\min_k \hat{z}_k(\mu_k^*)$. Using this equality and also
(\ref{eq:zk}), (\ref{eq:zk hat}), the proof is completed.
\end{proof}

Using Theorem \ref{theorem:TDMA}, the algorithm for Problem
\ref{prob:W-SPmin TDMA} can be obtained as follows: In each
iteration, the algorithm updates the dual variables $\mv{\mu}$ such
that $\hat{z}_1(\mu_1)=\cdots=\hat{z}_K(\mu_K)=c$. It then checks
whether the obtained rates can support more than the target rates,
and increases $c$ if they do or decreases it otherwise in the next
iteration, until the rate targets are exactly met and $c$ converges
to $c^*$. The details for the proposed algorithm are presented
below.
\begin{algorithm}\label{algorithm:minpower MAC TDMA}
\
\
\begin{itemize}
\item {\bf Given} $c_{\min}\leq c^* \leq c_{\max}$.
\item {\bf Repeat}
\begin{itemize}
\item [1.] $c\leftarrow \frac{1}{2}(c_{\min}+c_{\max})$.
\item [2.] For each $k$, obtain the optimal solutions $\mv{W}'_k$ and $\mu'_k$ such
that $\hat{z}_k(\mu'_k)=z_k(\mv{W}'_k,\mu'_k)=c$.\footnote{For each
given $\mu_k$, $\hat{z}_k(\mu_k)$ can be obtained by minimizing
$z_k(\mv{W}_k,\mu_k)$ over $\mv{W}_k$ as in (\ref{eq:zk hat}) by
means of a convex optimization method, e.g., the interior-point
method \cite{Boydbook}. The quantity $\mu'_k$ for which
$\hat{z}_k(\mu'_k)=c$ can then be obtained by a bisection search
over $\mu_k$ using the fact that $\hat{z}_k(\mu_k)$ is a decreasing
function of $\mu_k$.} Do the above for $k=1,\ldots,K$.
\item [3.] Compute $\tau'_k$ such that $R_k = \tau_k' \mathbb{E}_{\nu}\left[\frac{1}{2} \log \left|
\mv{H}_k(\nu)\mv{W}'_k\mv{H}_k^{\dag}(\nu)+\mv{I}\right|\right]$ for
$k=1,\ldots,K$.
\item[4.] If $\sum_{k=1}^{K}\tau'_k<1$, $c_{\min}\leftarrow c$;
otherwise $c_{\max}\leftarrow c$.
\end{itemize}
\item {\bf Until} $c_{\max}-c_{\min}<\delta$ where $\delta$ is a small positive constant that controls the algorithm accuracy.
\end{itemize}
\end{algorithm}
Since $\hat{z}_k(\mu_k^*)\leq 0, \forall k$, it follows that
$c^*\leq 0$. Thus, we can take $c_{\max}=0$. Similar as Appendix
\ref{appendix:initial ellipsoid}, we can obtain the upper bounds
$\mu_k^{(0)}$'s on the optimal dual solutions $\mu_k^*$'s. From
(\ref{eq:Lagrange power min TDMA}) and using the fact that
$g(\mv{\mu}^*)\geq 0$, it is easy to show that $c^*\geq
-\sum_{k=1}^{K}\mu_k^*R_k$. Thus, we can take $c_{\min}=
-\sum_{k=1}^{K}\mu_k^{(0)}R_k$. At last, the convergence of the
above algorithm as well as the uniqueness of the obtained solutions
are ensured by the uniqueness of $c^*$ in Theorem
\ref{theorem:TDMA}.

\section{Numerical Results}\label{sec:simulations}
\begin{figure}
\begin{center}
\scalebox{1.0}{\includegraphics{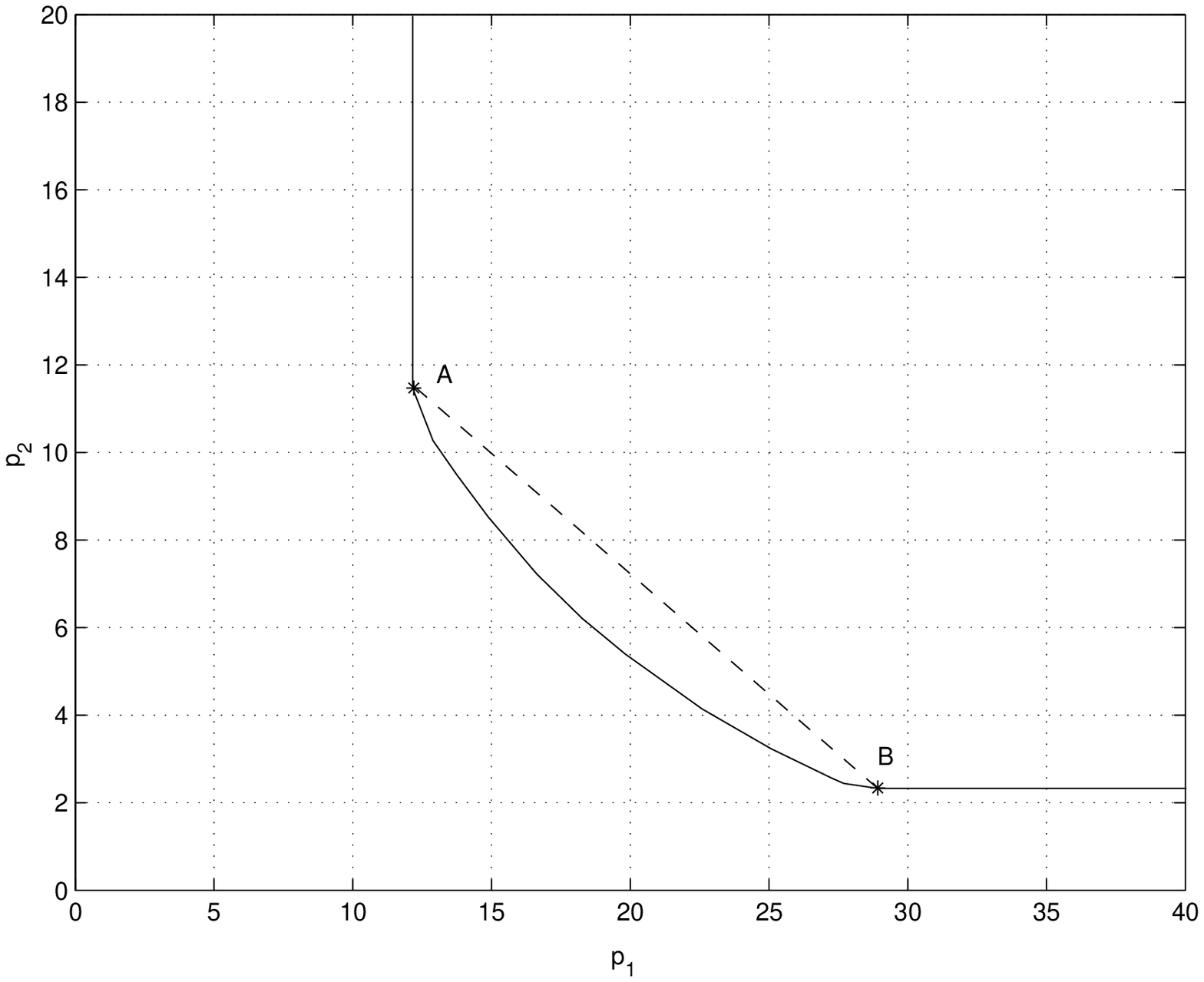}}
\end{center}
\caption{Power region for a two-user transmit-correlated fading
MIMO-MAC under SDMA with $t_1=t_2=2$, $r=2$, and the target rate,
$\mv{R} = [2~1]^T$ nats/sec/Hz.}\label{fig:power region SDMA}
\end{figure}

This section presents the power region for a fading MIMO-MAC with
$r=2$ receive antennas and $K=2$ MTs each equipped with $t_1=t_2=2$
transmit antennas. It is assumed that the receive antennas at the BS
are sufficiently separated that they experience independent fading,
while the fading levels are correlated across the transmit antennas
because of their realistic size limitations. Under this assumption,
the employed channel model for MT $k$ is given by $\mv{H}_k (\nu) =
\mv{H}_w (\nu) \mv{Q}_{k}^{1/2}$ for $k=1,2$, where
$\mv{Q}_{k}\in\mathbb{C}^{t_k \times t_k}$ denotes the transmit
antenna correlation matrix for MT $k$ and is assumed to be constant
over all fading states of $\nu$. $\mv{H}_w (\nu) \in \mathbb{C}^{r
\times t_k}$ denotes the Rayleigh-fading channel matrix that is
independent across two MTs and across all fading states, and has
independent entries distributed as $\mathcal{CN}(0,1)$. Similar as
the proof given in \cite{Syed01} and \cite{Ulukus05}, it can be
shown that the expressions in (\ref{eq:mac polymatroid}) and
(\ref{eq:zk}), for SDMA and TDMA, respectively, are both minimized
when the transmit signal covariance matrix, $\mv{S}_k$, has the same
set of eigenvectors as $\mv{Q}_{k}$, $k=1,2$, i.e., if $\mv{A}_k
\mv{\Lambda}_k \mv{A}_k^{\dagger}$ is the eigenvalue decomposition
of $\mv{Q}_{k}$, the optimal $\mv{S}_k$ then takes the form of
$\mv{A}_k \mv{\Sigma}_k \mv{A}_k^{\dagger}$ for some diagonal matrix
$\mv{\Sigma}_k$. This observation reduces the number of (real)
variables from $t_k^2$ in $\mv{S}_k$ to $t_k$ in $\mv{\Sigma}_k$ for
MT $k$ and in turn reduces the total algorithm
complexity.\footnote{The proposed algorithms work for all kinds of
CDI, e.g., with constant channel mean matrix, constant channel
covariance matrix, or combinations of them in all general forms.
However, for most of these cases, a similar variable-reduction like
in this numerical example is not possible.} Monte-Carlo simulation
with 5000 independent realizations of the random channels is used to
approximate the actual expectation over the fading states. The
simulation assumes that the target rate is $\mv{R} = [2~1]^T$
nats/sec/Hz for two MTs, and
\begin{eqnarray*}
\mv{Q}_{1} = \left[%
\begin{array}{cc}
  1 & 0.4 \\
  0.4 & 1 \\
\end{array}%
\right],             \ \ \ \ \ \ \ \ \ \ \
\mv{Q}_{2} = \left[%
\begin{array}{cc}
  1 & 0.5 \\
  0.5 & 1 \\
\end{array}%
\right].
\end{eqnarray*}
Fig. \ref{fig:power region SDMA} and Fig. \ref{fig:power region
TDMA} show the power region for this fading MAC under SDMA and TDMA
obtained by solving Problem \ref{prob:W-SPmin SDMA} and Problem
\ref{prob:W-SPmin TDMA}, respectively.

For the SDMA case as is shown in Fig. \ref{fig:power region SDMA},
there are two corner points on the boundary of the power region,
denoted by point A and B, which can be obtained by the greedy
algorithm described in Section \ref{subsec:weighted sum power}.
Recall that in this greedy algorithm, the BS first picks one
possible decoding order for the MTs, and then starting from the last
decoded MT, it minimizes the power required to maintain the target
rate for each MT while considering the MTs that have yet been
decoded as interference. In this figure, the power-pair A
corresponds to the decoding order $2\rightarrow1$, while the
power-pair B does for the reversed decoding order. The greedy
algorithm achieves the optimal power-pairs for the W-SPmin problem
under some weight vectors, e.g., point A for $\lambda_1\gg\lambda_2$
and point B for $\lambda_2\gg\lambda_1$, but might be suboptimal for
weight vectors other than these extreme choices. For example, the
minimum value of $0.4 p_1+ 0.6 p_2$ to support the target rate
$[2~1]^T$ is 11.5 while the greedy algorithm leads to 12.8 and 13.3
units of power for power-pair A and B, respectively. Moreover, the
boundary curve of the power region is not attainable by simply
time-sharing these two corner points, as shown by the dashed line in
Fig.\ref{fig:power region SDMA}.

\begin{figure}
\begin{center}
\scalebox{1.0}{\includegraphics{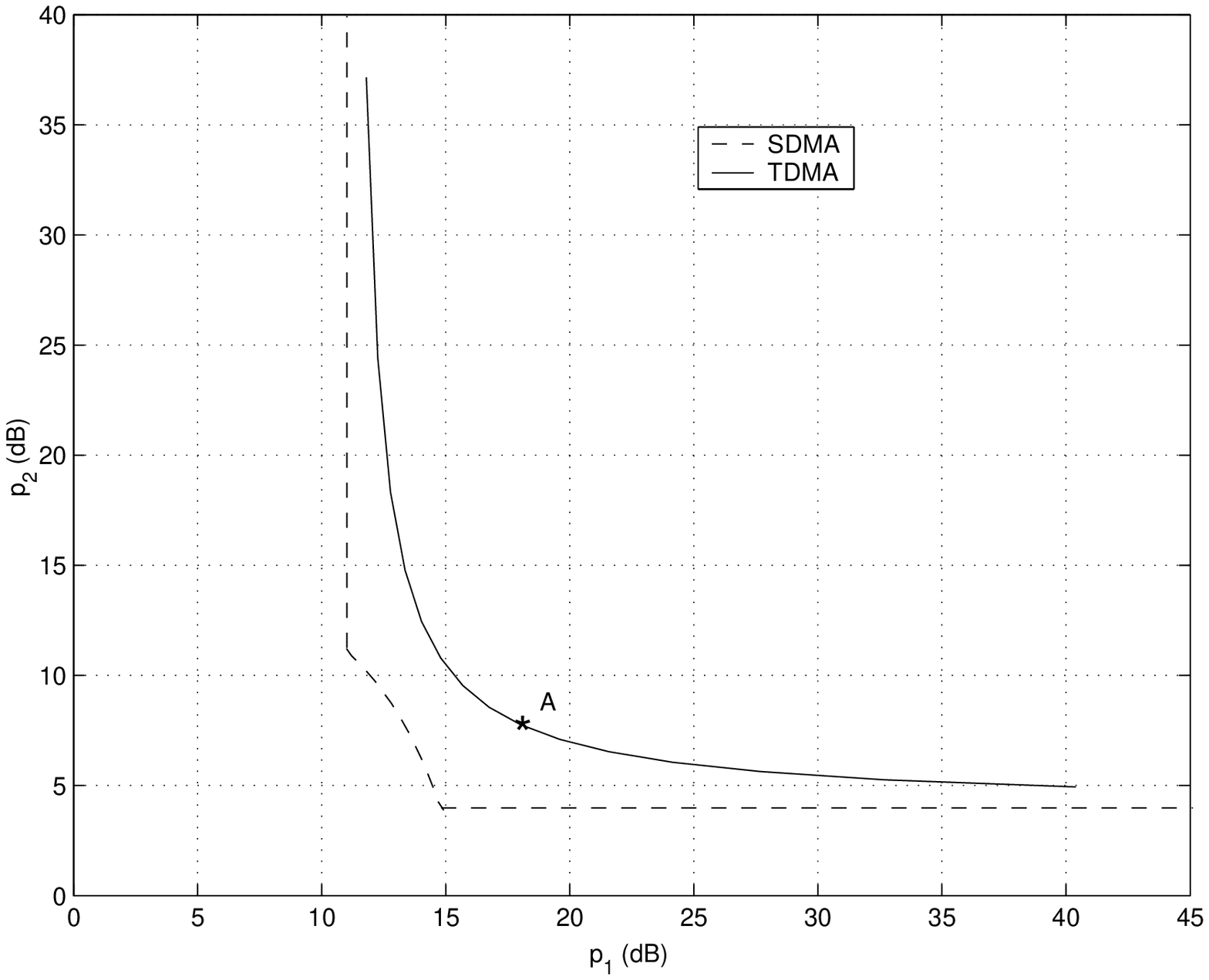}}
\end{center}
\caption{Power region for a two-user transmit-correlated fading
MIMO-MAC under SDMA and TDMA with $t_1=t_2=2$, $r=2$, and the target
rate, $\mv{R} = [2~1]^T$ nats/sec/Hz.}\label{fig:power region TDMA}
\end{figure}

The power region under TDMA for this MAC is shown in Fig.
\ref{fig:power region TDMA}. For comparison, the power region under
SDMA is also included in this figure. This figure uses the log scale
for the powers and, hence, the power region under SDMA looks as a
non-convex set. The power savings achieved by SDMA are observed to
be substantial compared to TDMA, even in this case where the number
of transmit antennas at each MT is equal to that of the receive
antennas at the BS, i.e., both SDMA and TDMA have the same number of
degrees of transmission freedom in the spatial domain, which is two
in this case. The power-pairs on the boundary of the power region
under TDMA correspond to different time-slot durations, $\tau_k$,
$k=1,2$, assigned to each MT. The power-pair A shown in Fig.
\ref{fig:power region TDMA} is achieved by assigning equal duration
of time slot for both MTs, i.e., $\tau_1=\tau_2=0.5$ as in the
conventional TDMA. Clearly, this power-pair is optimal for the
W-SPmin problem under a unique weight vector $\mv{\lambda}$, and for
the SPmin-PPC problem under a unique power-profile vector
$\mv{\alpha}$, but it is suboptimal in all the other cases. For
example, the minimum sum-power $p_1+p_2$ to achieve the target rate
$[2~1]^T$ is 42 units of power for $\tau_1=0.66$ and $\tau_2=0.34$,
as compared to 69 units of power for $\tau_1=\tau_2=0.5$.

\section{Conclusions}\label{sec:conclusions}

This paper characterizes the power region for the fading MIMO-MAC.
Motivated by a general relationship between the power region and the
corresponding capacity region, the Lagrange primal-dual approach is
employed to characterize all pareto optimal power-tuples on the
boundary of the power region. These optimal power-tuples provide
different power tradeoff among the MTs and also ensure the fairness
of power consumption among them. The algorithms developed in this
paper can be used in the wireless cellular network for the BS to
control the transmit powers from the MTs in the uplink transmission.
Two multiple access techniques, namely, SDMA and TDMA, are
considered in this paper. It is observed that substantial power
savings can be obtained by using SDMA compared to TDMA. This
observation provides an important information-theoretic guidance for
practical system designs, i.e., if the complexity for implementing
the optimal SDMA can be tailed for, an enormous capacity gain is
still possible over the conventional TDMA-based network. The
multiuser transmit-covariance feedback scheme studied in this paper
optimizes the transmit covariance matrices of all the MTs based on
their long-term CDI. Hence, this scheme reduces significantly the
feedback complexity compared to other feedback schemes based on the
instantaneous channel realizations. As a result, this scheme is
practically suitable for wireless channels that exhibit some
consistent long-term channel statistics. The results obtained in
this paper can provide insightful guidelines to many applications in
wireless networks including resource allocation, partial channel
feedback, and multiuser space-time code design.

\appendices

\section{Initial Ellipsoid for Algorithm \ref{algorithm:minpower MAC
SDMA}} \label{appendix:initial ellipsoid}

In the appendix, one possible method to obtain the initial ellipsoid
$\mathcal{E}[0]$ for Algorithm \ref{algorithm:minpower MAC SDMA} is
presented. First, we obtain an upper bound $\mu_j^{(0)}$ on
$\mu_j^*$ for any given $j \in \{1,\ldots,K\}$. Let
$\{\mv{S}_k^{(j)}\} \in \mathcal{D}$ be any set of transmit
covariance matricies that achieve $\{r_k^{(j)}\}$ given by
\begin{eqnarray}
 r_k^{(j)}=\left\{%
\begin{array}{c}
  R_k^* \ \ \ \ k \neq j \\
  R_k^* +1 \ k = j. \\
\end{array}%
\right.
\end{eqnarray}
From the definition of the dual function given by (\ref{eq:Lagrange
dual power min SDMA}), we have
\begin{eqnarray*}
g(\mv{\mu}^*)\leq
\mathcal{L}(\{\mv{S}_k^{(j)}\},\{r_k^{(j)}\},\mv{\mu}^*) =
\sum_{k=1}^{K}\lambda_k \mathtt{Tr}\left(\mv{S}_k^{(j)}\right) -
\mu_j^*.
\end{eqnarray*}
Since $g(\mv{\mu}^*)\geq0$, it follows that
\begin{equation}
\mu_j^* \leq \sum_{k=1}^{K}\lambda_k
\mathtt{Tr}\left(\mv{S}_k^{(j)}\right).
\end{equation}
Thus, $\mu_j^{(0)}=\sum_{k=1}^{K}\lambda_k
\mathtt{Tr}\left(\mv{S}_k^{(j)}\right)$. Similar upper bounds can be
found for all other $j$. Next, $\mathcal{E}[0]$ can be chosen to
cover the hyper-cube in $\mathbb{R}^K$ specified by $\mu_j^{(0)}$'s.

\section{Proof of Theorem \ref{theorem:uniqueness of Sk}} \label{appendix:proof Sk}

This Appendix proves the uniqueness of the solutions for the optimal
transmit covariance matrices $\{\mv{S}_k^*\}$ in Problem
\ref{prob:W-SPmin SDMA}. Without loss of generality, it is assumed
that $\mu_1^*\geq\cdots\geq\mu_K^*>\mu_{K+1}^*=0$. If
$\{\mv{S}_k^{(1)}\}$ and $\{\mv{S}_k^{(2)}\}$ are two sets of
optimal solutions for Problem \ref{prob:W-SPmin SDMA}, from
(\ref{eq:Lagrange power min SDMA}) and by using Theorem
\ref{theorem:polymatroid}, it follows that
\begin{eqnarray}\label{eq:two optimal Sk}
q^* = \sum_{k=1}^K \lambda_k
\mathtt{Tr}(\mv{S}_k^{(j)})+\sum_{k=1}^K\mu_k^*R_k-
\sum_{k=1}^{K}\left(\mu_k^* - \mu_{k+1}^*\right)
\mathbb{E}_{\nu}\left[
\frac{1}{2}\log\left|\sum_{i=1}^{k}\mv{H}_{i}(\nu)\mv{S}_{i}^{(j)}\mv{H}_{i}(\nu)^{\dagger}
+\mv{I}\right|\right],
\end{eqnarray}
for $j=1,2$. Since the problem at hand is convex, for any $\beta
\in [0,1]$, $\mv{S}_k = \beta
\mv{S}_k^{(1)}+\bar{\beta}\mv{S}_k^{(2)}$ is also an optimal
solution to satisfy (\ref{eq:two optimal Sk}), where
$\bar{\beta}=1-\beta$. This fact together with the concavity of
the $\log|\cdot|$ function implies that
\begin{eqnarray}
\mathbb{E}_{\nu}\left[\log\left|\beta\mv{A}^{(1)}(\nu)+\bar{\beta}\mv{A}^{(2)}(\nu)\right|-
\beta\log\left|\mv{A}^{(1)}(\nu)\right|-\bar{\beta}\log\left|\mv{A}^{(2)}(\nu)\right|
\right]=0,
\end{eqnarray}
where $\mv{A}^{(j)}(\nu)\triangleq
\sum_{k=1}^{K}\mv{H}_{k}(\nu)\mv{S}_{k}^{(j)}\mv{H}_{k}(\nu)^{\dagger}
+\mv{I}$. Let $f(\beta)$ denote the function on the LHS of the
above equation, then $f(\beta)=0$, for all $0\leq\beta\leq 1$.
Because $f(\beta)$ is twice continuously differentiable, both of
its first and second derivatives must vanish, i.e.,
\begin{eqnarray}
\frac{d^2f(\beta)}{d\beta^2}=-\mathbb{E}_{\nu}\left[
\mathtt{Tr}\left(\left(\left(\mv{A}^{(1)}(\nu)-\mv{A}^{(2)}(\nu)\right)\left(\beta\mv{A}^{(1)}(\nu)
+
\bar{\beta}\mv{A}^{(2)}(\nu)\right)^{-1}\right)^2\right)\right]=0,
\ \ \forall \beta.
\end{eqnarray}
For every $\nu$, the matrix in $\mathtt{Tr}(\cdot)$ of the above
equation is a positive semi-definite matrix and, hence, it has a
nonnegative trace. Since the expectation of a nonnegative random
variable is zero, it must be zero {\it a.s.}, or
$\mv{A}^{(1)}(\nu)=\mv{A}^{(2)}(\nu)$ {\it a.s.}, which implies that
$\mv{S}_{k}^{(1)}=\mv{S}_{k}^{(2)}$.

\newpage

\begin{figure}
\psfrag{a}{$p_1$} \psfrag{b}{$p_2$} \psfrag{c}{$(p_1^*,p_2^*)$}
\psfrag{d}{$\mv{\lambda}^T\mv{p}$} \psfrag{e}{$r_1$}
\psfrag{f}{$r_2$} \psfrag{g}{$(R_1,R_2)$}
\psfrag{h}{$\mv{\rho}^T\mv{r}$}
\psfrag{i}{$\frac{\lambda_1}{h_1}>\frac{\lambda_2}{h_2}$}
\psfrag{j}{Decoding Order $1\rightarrow 2$}\psfrag{k}{Decoding Order
$2\rightarrow 1$}
\begin{center}
\scalebox{1.0}{\includegraphics*[41pt,544pt][471pt,794pt]{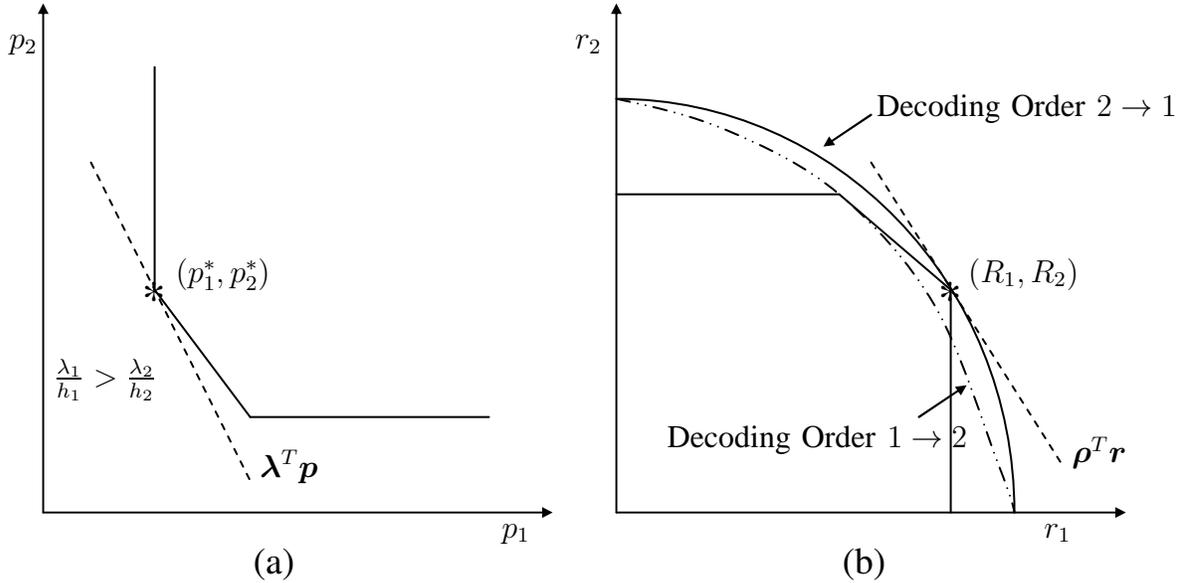}}
\end{center}
\caption{The relationship between each boundary point of the power
region and its corresponding capacity region for a two-user
deterministic SISO-MAC under SDMA. (a) Power region under the rate
constraint $(R_1,R_2)$, where $(p_1^*,p_2^*)$ is on its boundary and
achieves the minimum weighted sum-power for the weights $\lambda_1$
and $\lambda_2$, and the associated decoding order is $2\rightarrow
1$ since $\frac{\lambda_1}{h_1}>\frac{\lambda_2}{h_2}$; (b)
Corresponding capacity region of $(p_1^*, p_2^*)$ under the weighted
sum-power constraint $\lambda_1p_1+\lambda_2p_2\leq p^*$, where
$p^*=\lambda_1p_1^*+\lambda_2p_2^*$.} \label{fig:power capacity
SISO}
\end{figure}

\end{document}